\documentclass[onecolumn,notitlepage,superscriptaddress,11pt,tightenlines,nofootinbib]{revtex4}    
\makeatletter
\usepackage[ colorlinks = true,
             linkcolor = blue,
             urlcolor  = blue,
             citecolor = red,
             anchorcolor = green,
]{hyperref}
\usepackage{comment}
\usepackage{tikz}
\usepackage{pgfbaseplot}
\usepackage{overpic}
\usepackage{graphicx,amsfonts,amssymb,amsmath,amsthm}
\usepackage{xcolor}
\usepackage{braket,enumerate}
\usepackage{mathtools}
\mathtoolsset{showonlyrefs}
\newcommand{\eps}{\varepsilon}
\newcommand{\E}{\mathbb{E}}

\newcommand{\M}{{\mathcal{M}}}

\newtheorem{thm}{Theorem}
\newtheorem{corollary}[thm]{Corollary}
\newtheorem{lemma}[thm]{Lemma}

\newtheorem{assumption}[thm]{Assumption}

\newtheorem{definition}[thm]{Definition}
\theoremstyle{remark}

\newtheorem{example}[thm]{Example}

\newtheorem{remark}[thm]{Remark}

\newtheorem{tab}{Table}

\DeclareMathOperator{\Tr}{Tr}
\newcommand{\id}{I}
\renewcommand{\P}{\mathbb{P}}

\begin{document}
\title{\Large Optimal Adaptive Strategies for  \\ Sequential Quantum Hypothesis Testing}
 \author{Yonglong~Li}
  \email[]{elelong@nus.edu.sg}
  \affiliation{Department of Electrical and Computer Engineering, National University of Singapore}

\author{Vincent Y.~F.~Tan}
  \email[]{vtan@nus.edu.sg}
  \affiliation{Department of Mathematics, \\ Department of Electrical and Computer Engineering, \\ Institute of Operations Research and Analytics, National University of Singapore}
  
\author{Marco Tomamichel}                   
 \email[]{marco.tomamichel@nus.edu.sg}
 \affiliation{Department of Electrical and Computer Engineering, and \\Centre for Quantum Technologies, National University of Singapore}

\begin{abstract}
We consider sequential hypothesis testing between two quantum states using adaptive and non-adaptive strategies. In this setting, samples of an unknown state are requested sequentially and a decision to either continue or to accept one of the two hypotheses is made after each test. Under the constraint that the number of samples is bounded, either in expectation or with high probability, we exhibit adaptive strategies that minimize both types of misidentification errors. Namely, we show that these errors decrease exponentially (in the stopping time) with decay rates given by the measured relative entropies between the two states. Moreover, if we allow joint measurements on multiple samples, the rates are increased to the respective quantum relative entropies. We also fully characterize the achievable error exponents for non-adaptive strategies and provide numerical evidence showing that adaptive measurements are necessary to achieve our bounds.
\end{abstract}
\maketitle

\section{Introduction}

We consider the binary quantum hypothesis testing problem~\cite{helstrom67}, where an unknown quantum state~$\rho$ is either given by a density operator $\rho_0$ or $\rho_1$, and we are tasked to devise a measurement strategy that efficiently determines which of these two hypotheses is true. This problem lies at the core of quantum physics since it provides a rigorous theoretical framework for one of physics' most fundamental tasks: determining which mathematical model best describes a physical system.  Beyond that binary quantum hypothesis testing has various applications in quantum information theory, for example in quantum channel coding problems (see, e.g.,~\cite{hayashi03,wang10}).

\begin{figure}
\centering
\begin{overpic}[width=.8\textwidth]{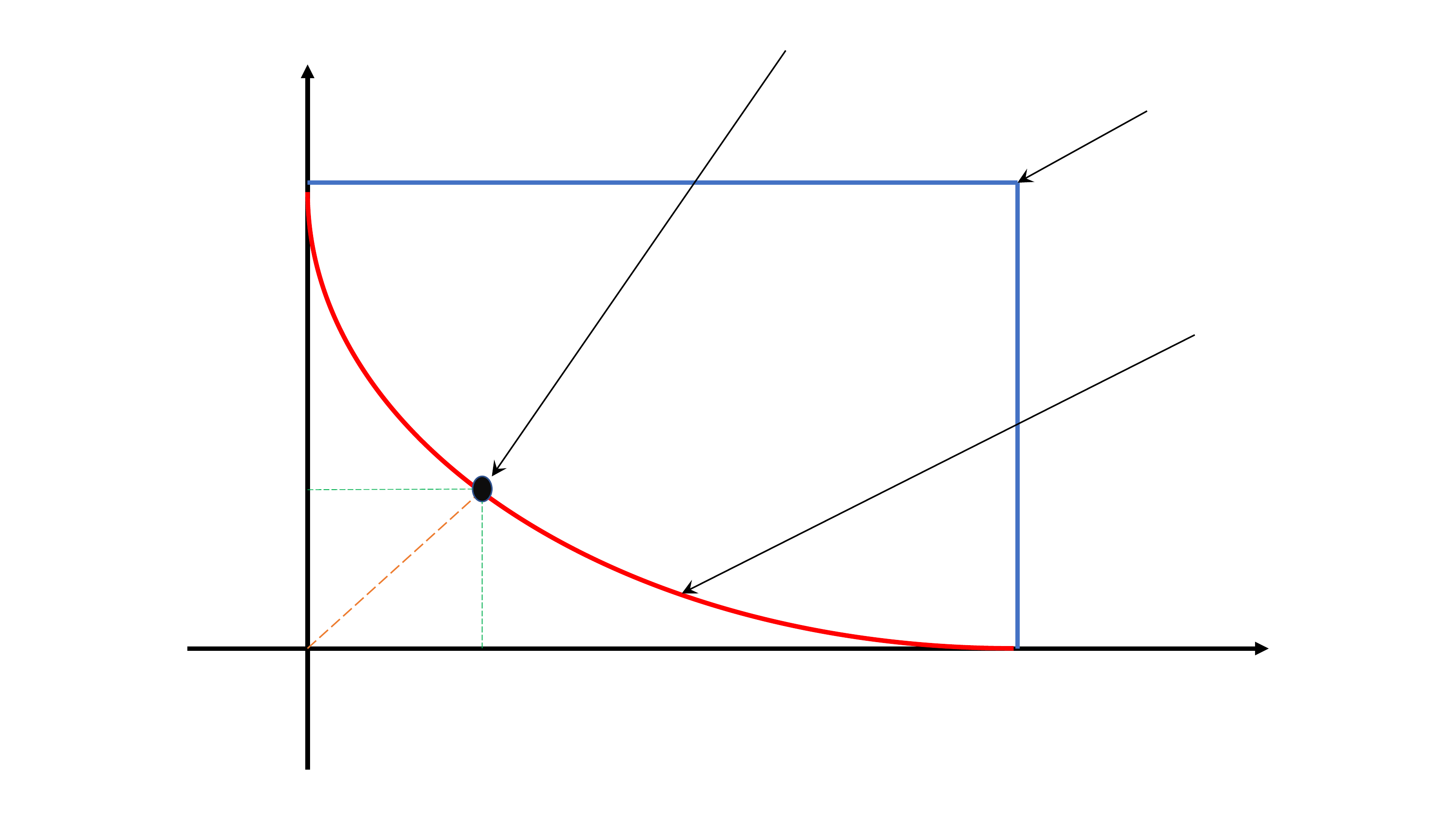}
\put(67,8){$D(\rho_1\|\rho_0)$}
\put(27,8){$C(\rho_0, \rho_1)$}
\put(9,22){$C(\rho_0, \rho_1)$}
\put(80,16){$\displaystyle\lim_{n\to\infty} \frac{1}{n}\log \frac{1}{\alpha_n}$}
\put(41,54){Chernoff exponent}
\put(73,35){Tradeoff given by Hoeffding}
\put(67,50){Tradeoff given by sequential tests}
\put(9,43){$D(\rho_0\|\rho_1)$}
\put(1,53){$\displaystyle\lim_{n\to\infty} \frac{1}{n}\log\frac{1}{ \beta_n}$}
\put(19,9){$0$}
\end{overpic}
\caption{Schematic of the optimal trade-off between the exponential decay rates for the error of the first and second kind}
\label{fig:hoeffding}
\end{figure}
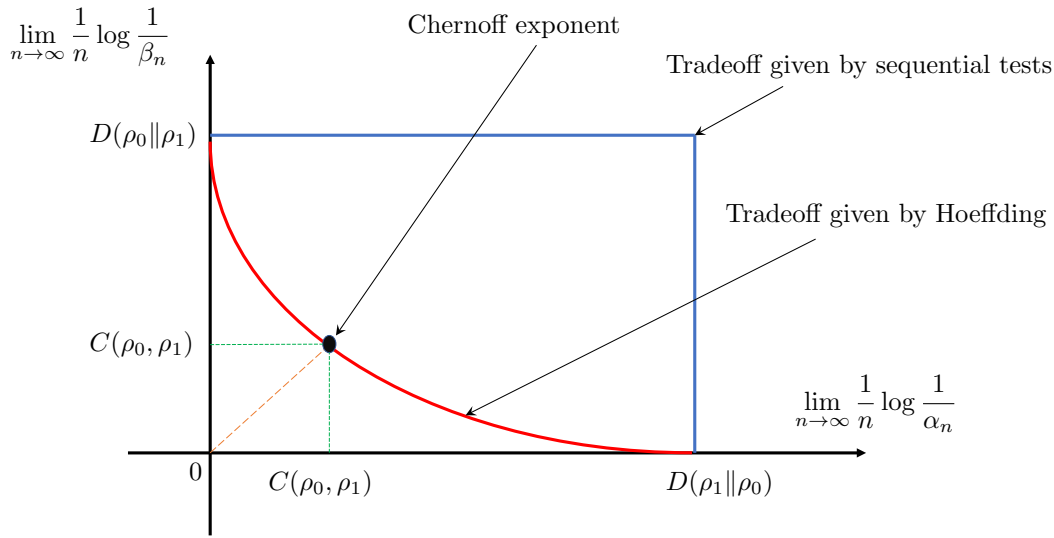

In the usual \emph{fixed-length} setup, we are given $n$ samples of the state $\rho$ and can perform a test, i.e., a measurement $\{ \Lambda_n , \id_n - \Lambda_n\}$, on the joint quantum system 
to determine if the first or second hypothesis is correct. The two types of misidentification errors, of the first and second kind, are then respectively given by
\begin{align}
	\alpha_n = \Tr \left( \rho_0^{\otimes n} (\id_n - \Lambda_n) \right) \qquad \textrm{and} \qquad \beta_n = \Tr \left( \rho_1^{\otimes n} \Lambda_n \right) \,.
\end{align}
Our goal subsequently is to find tests that minimize these two error probabilities, or more specifically, to find the optimal trade-off between them. Many of the early results on this topics are reviewed in~\cite{audenaert07-3}. The quantum generalization of Stein's lemma~\cite{Petz1991,Nagaoka2000} establishes that when the first kind of error is upper bounded by a constant, the error of the second kind decays exponentially with Stein's exponent given by the \emph{quantum relative entropy}, $D(\rho_0 \| \rho_1)$. On the other hand, if both errors decrease exponentially, the optimal trade-off between the decay rates is governed by the quantum Hoeffding bound~\cite{Hayashi2004,Nagaoka2006,Hayashi2007}. The symmetric case when both exponents are required to be the same is covered by the quantum Chernoff exponent~\cite{Audenaert2007,Nussbaum2009}, $C(\rho_0, \rho_1)$. These results are summarized in Figure~\ref{fig:hoeffding}. Beyond this, second-order refinements to Stein's exponent were derived in~\cite{tomamichel12, li12} and the moderate deviation regime where one error probability decreases sub-exponentially has been analysed in~\cite{chubb17,cheng17}. We can further impose the restriction that each sample is measured instantaneously, which reduces the extremal Stein exponents to the \emph{measured relative entropy}, $D_{\mathcal{M}}(\rho_0 \| \rho_1)$. Under these restrictions, \emph{adaptive} strategies, where the choice of subsequent measurements may depend on previous observations, become meaningful. They have been investigated in~\cite{Hayashi2009} and do not yield any improvements over non-adaptive strategies.\footnote{On the other hand, it is worth noting that adaptive strategies bring an advantage in quantum channel discrimination~\cite{salek20} in the fixed-length setup.}

What these results and Figure~\ref{fig:hoeffding} show is that for a {\em fixed-length} binary quantum hypothesis testing problem, there exists a fundamental trade-off between the two kinds of decay rates;  they cannot assume the extremal values $D(\rho_1\| \rho_0)$ and $D(\rho_0\| \rho_1)$ simultaneously. 
Is it then possible to go beyond the Hoeffding and Chernoff bounds? In a recent paper~\cite{quantumSHT}, an affirmative answer to this question was given in the setting of {\em sequential} quantum hypothesis testing. The sequential approach to quantum hypothesis testing was first explored in~\cite{slussarenko17}. 

In the statistical literature, sequential methods were first proposed much earlier in~\cite{Wald1945} for the classical hypothesis testing problem. Instead of fixing the sample size before performing the hypothesis test, sequential methods allow the sample size to be a {\em random variable}. In particular, at  each time the experimenter will request a new sample if the current set of samples does not give the experimenter sufficient confidence to make a decision that meets the error criteria. We then require that the sample size is bounded either in expectation or with high probability. Somewhat surprisingly, sequential methods can decouple the two kinds of error probabilities and allow  the experimenter to control both. When the length of the hypothesis test (i.e., the number of observed samples) is allowed to be a random variable whose {\em expectation} is bounded by $n$ for the binary hypothesis testing problem between two probability distributions $P_0$ and $P_1$, it was shown in~\cite{WaldWolf} that   there exists a sequence of tests---namely sequential probability ratio tests (SPRTs)---such that the exponents of the errors of the first and second kind {\em simultaneously} assume the extremal values  $D(P_1\| P_0)$ and $D(P_0\| P_1)$. This significantly improves the classical Hoeffding bound of the error exponents~\cite{Hoeffding1965, Blahut} where if one error exponent assumes its extremal value---the relative entropy---the other necessarily vanishes.

\begin{figure}
\centering
\begin{overpic}[width=.8\textwidth]{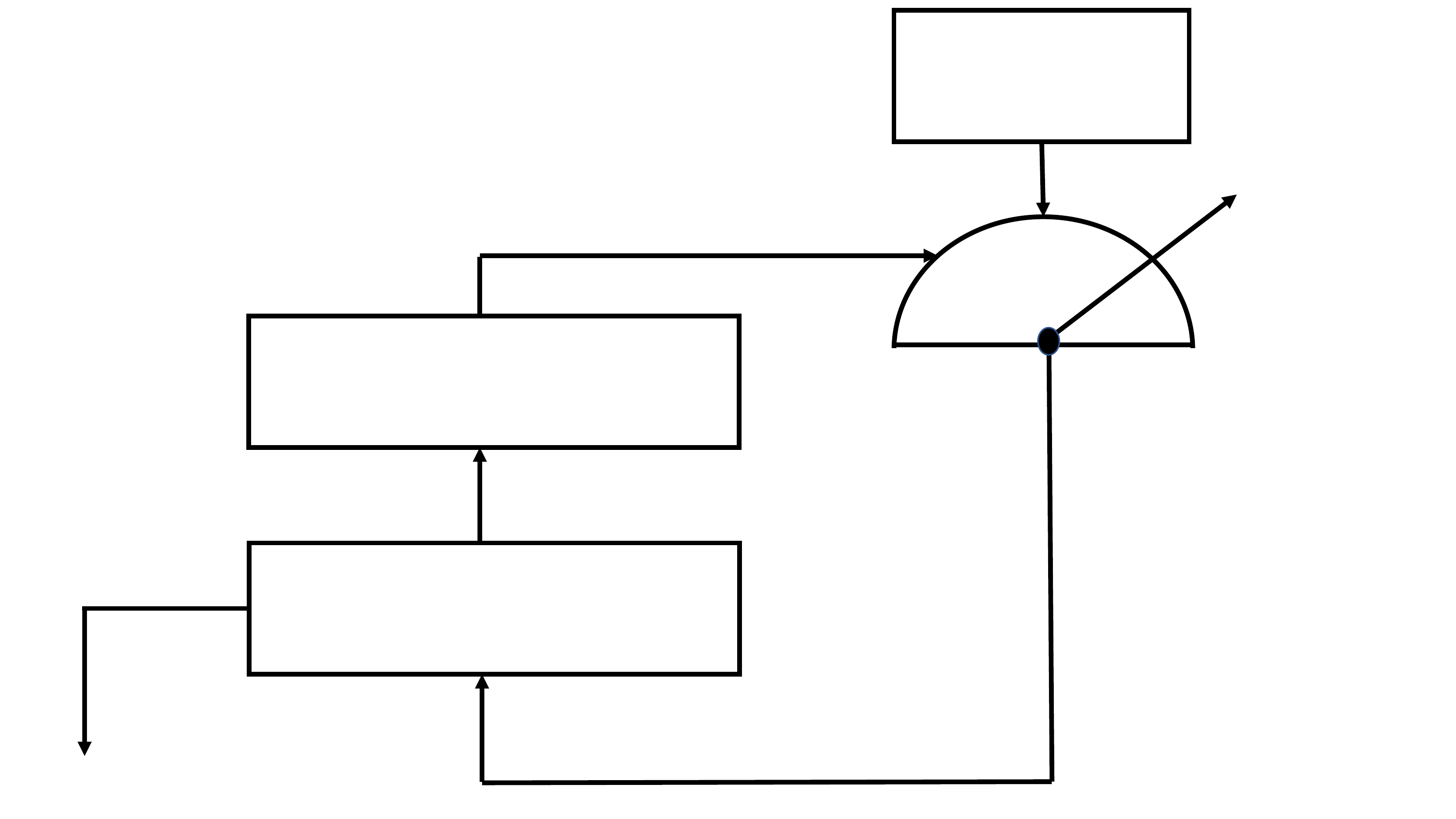}
\put(48,41){$m_{k+1}$}
\put(67,50){Oracle}
\put(73,43){$\rho_0/\rho_1$}
\put(22,29){$\mu_{k+1}(\mathrm{d}m_{k+1 }|m_1^k,x_1^k)$}
\put(34,21){$*$}
\put(27,14){$d_{k}(m_1^k,x_1^k)$}
\put(52,4){$x_k$}
\put(3,2.5){Stop}
\put(-3,10){$0$ or $1$}
\end{overpic}
\caption{The structure of a general adaptive sequential hypothesis testing protocol.}
\label{fig:adaptive}
\end{figure}
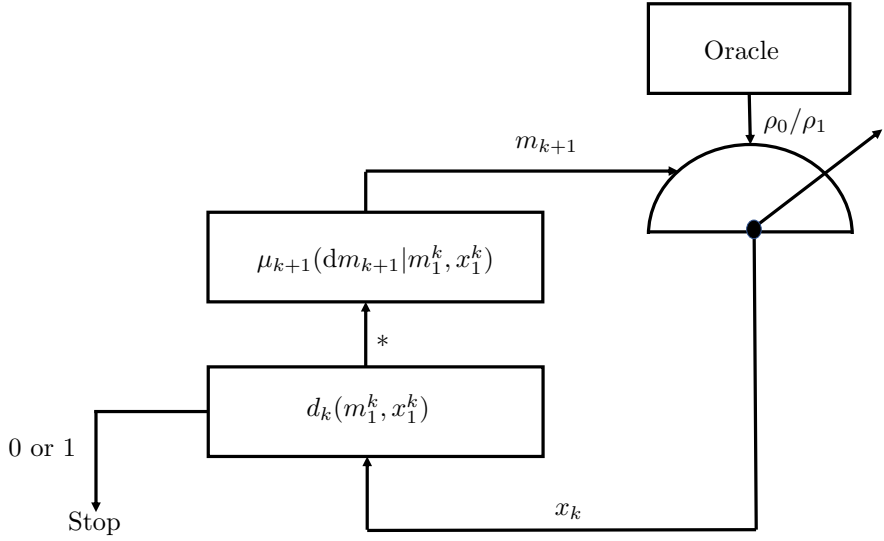

As seen in Figure~\ref{fig:adaptive} in the sequential setting we study in this paper we are required to make decisions after observing each individual sample, and thus joint measurements on all $n$ samples are generally not optimal. In contrast, adaptive measurement strategies where the measurement at time $k$ can be based on measurements and outcomes before time $k$ are now an intriguing prospect. 
Similar control strategies have been used in the classical setting. In~\cite{PV10}, the region of achievable error exponent pairs for  sequential binary hypothesis testing with feedback, wherein the two hypotheses are characterized by discrete memoryless channels $W$ and $V$, was characterized. In addition to being able to access feedback from the receiver, the sender is also able to adaptively control the channel inputs. It was shown that the control strategy used in~\cite{PV10} is asymptotically optimal in a certain Bayesian setting studied in~\cite{NJ13}. 

In this paper we show that in sequential quantum hypothesis testing, where the number of samples is a random variable, adaptive strategies allow us to attain the corner point in Figure~\ref{fig:hoeffding}. As such, one can enjoy the {\em best of both worlds} in terms of the extremal decay rates $D(\rho_1\| \rho_0)$ and $D(\rho_0\| \rho_1)$. Our main results are summarized as follows:

\begin{enumerate}
 \item Given that the number of samples is bounded by $n$ either in expectation or with probability exceeding $1-\eps$ for some $0<\eps<1$, we show that there exists a sequence of adaptive sequential measurement strategies that measures each sample instantaneously achieves the decay rates $D_{\mathcal{M}}(\rho_1\| \rho_0)$ and $D_{\mathcal{M}}(\rho_0\| \rho_1)$ for the errors of the first and second kind, respectively. This improves on the results in~\cite{quantumSHT}, where this rate pair can only be achieved if there exists a single measurement that achieves both measured relative entropies, and no matching converse is given. This is presented as Theorems~\ref{expectation} and~\ref{probabilistic} in Section~\ref{sec:adapt}.

 \item Under the same constraint on the number of samples but with the additional freedom that samples can be stored and measured jointly, we show that the decay rates $D(\rho_1\| \rho_0)$ and $D(\rho_0\| \rho_1)$ for the errors of the first and second kind can be achieved. This resolves a problem left open in~\cite{quantumSHT}, where a converse showing that this rate pair is optimal under the expectation constraint was established, but no protocol achieving it was analyzed. 
We note that the converse result in~\cite{quantumSHT} covers the most general adaptive strategies involving a quantum memory, but we show that block-measurement strategies are sufficient to achieve it. This is presented as Theorem~\ref{ultimate:adaptive} in Section~\ref{sec:adapt}.
 
 \item We also provide a full characterization for non-adaptive sequential quantum hypothesis testing, where the same measurement is used for every sample. This is presented as Theorem~\ref{non-adaptive} in Section~\ref{sec:non-adapt}. Using this, in Section~\ref{sec:num}, we exhibit a numerical example where the decay rates cannot achieve the measured relative entropies using non-adaptive measurements. It thus shows that adaptive measurements are necessary to unlock the full power of sequential quantum hypothesis testing. 
\end{enumerate}

The reminder of the paper is structured as follows.
In Section~\ref{sec:model} we formally introduce the mathematical model for sequential quantum hypothesis testing. In Section~\ref{main}, for different testing strategies, we state the maximal achievable regions of error exponent pairs under different type of constraints on the number of copies of quantum states used in sequential quantum hypothesis tests. In Section~\ref{sec:preli}, we collect some tools used in the proof of our main theorems and prove the main theorems in Section~\ref{sec:proof}.

\section{Problem Setting}\label{sec:model}

\subsection{Notation}

In the following, let us fix $\mathbb{C}^d$ as the $d$-dimensional Hilbert space. A \emph{quantum state} is given by a positive semidefinite matrix with unit trace in $\mathbb{C}^{d\times d}$. We say that a quantum state has full support if all eigenvalues are strictly positive. A \emph{positive operator-valued measure} (POVM) is given by a finite set $\mathcal{X}$ and a collection of positive semidefinite matrices $m = \{ m(x) \}_{x \in \mathcal{X}}$ in $\mathbb{C}^{d \times d}$ such that $\sum_{x \in \mathcal{X}} m(x) = \id$, where $\id$ is the identity matrix in $\mathbb{C}^{d\times d}$. (Somewhat unconventionally, throughout this paper $M$ is used to denote a {\em random} POVM and $m$ is used to denote a {\em realization} of such a random POVM.) The probability of observing an outcome $x \in \mathcal{X}$ on a state $\rho$ is then given by Born's rule, $P_{\rho,m}(x) = \Tr [ \rho m(x) ]$. 
A \emph{projector-valued measure} (PVM) is a POVM that additionally satisfies $m(x) m(x) = m(x)$ for all $x \in \mathcal{X}$, i.e., all $m(x)$ are projectors. Rank-$1$ PVMs have the additional property that all projectors have only a single non-zero eigenvalue.
Let $\mathcal{M}_{\mathcal{X}}$ be the set of POVMs on $\mathbb{C}^{d\times d}$ that are indexed by $\mathcal{X}$. As $\mathcal{X}$ is finite, $\M_\mathcal{X}$ is a compact subset of $\mathbb{C}^{d\times d} \times \mathcal{X}$ equipped with the usual Euclidean topology. 
We will be looking at sequences of POVMs $m_1^k = (m_1, m_2, \ldots, m_k)$, and sequences of measurement outcomes $x^{k}_{1} = (x_1, x_2, \ldots, x_k)$.

\subsection{Sequential Tests}\label{subsec:sqht}

We consider now the binary quantum hypothesis testing problem where an unknown quantum state $\rho$ is either $\rho_0$ or $\rho_1$.
In this work we consider \emph{sequential tests} to determine which of the two hypotheses is in effect. A sequential quantum hypothesis test (SQHT) $\mathcal{S}=\big(\mathcal{X},\{\mu_{k},d_{k}\}_{k=1}^{\infty}\big)$, in its most general form, is given by (see also Figure~\ref{fig:adaptive}):
\begin{itemize}
	\item a finite set of measurement outcomes, $\mathcal{X}$;
	\item a sequence of conditional  probability measures to determine the next measurement, $\mu_k(\mathrm{d}m_k | x_{1}^{k-1}, m_{1}^{k-1})$ for every $k \in \mathbb{N}$;
	\item a sequence of $\{0,1,*\}$-valued decision functions $d_k(x_{1}^{k}, m_{1}^{k})$, for every $k \in \mathbb{N}$.
\end{itemize}
We remark that our results can be generalized to random decision functions $d_k(x_{1}^{k}, m_{1}^{k})$. However, for notational simplicity, we only consider deterministic decision functions.  

\newcommand{\cX}{\mathcal{X}}

At time $k \geq 1$, the experimenter chooses the POVM $M_k = m_k$ randomly according to the conditional probability measure $\mu_k(\cdot|x_{1}^{k-1},m_{1}^{k-1})$. After choosing the POVM, the experimenter then applies $m_k$ to the $k$-th sample of the underlying unknown state $\rho$ and obtains the outcome $X_k = x_k$ with probability $\Tr [ \rho m_k(x_{k}) ]$. Then based on the POVMs $m_1^k$ and the outcomes $x_1^k$, the experimenter chooses to either stop or continue the test. At each time $k$, the actions of the experimenter can be described by a $\{0,1,*\}$-valued function $d_k$. If $d_k=*$, the experimenter continues the test after applying $m_{k}$ to the underlying state and if  $d_k = i\in \{0,1\}$, the experimenter stops the test and declares $\rho_i$ to be the underlying state. Let $T$ be the first time that $d_{k}\not=*$. Thus the number of samples of the underlying state $\rho$ used during the test is $T$.  This testing protocol is depicted in Figure~\ref{fig:adaptive}. We point out two important features of our problem setting and test: 
\begin{itemize}
\item[(a)] The number of samples of the underlying state $\rho$ used during the test is not {\em fixed} but is a {\em random variable} denoted by $T$ in the sequel;
\item[(b)] The testing strategy is {\em adaptive} since the POVM used at time $k$ can depend on all the POVMs and outcomes before time $k$.
\end{itemize}
Throughout the rest of the paper, a SQHT will be denoted by $\mathcal{S}=\big(\cX,\{\mu_{k},d_{k}\}_{k=1}^{\infty},T\big)$ to emphasize the (random) number of samples $T$ used in a SQHT. When $\mu_k$ is a probability measure on~$\M_\mathcal{X}$ and in particular, $\mu_k$ does {\em not} depend on $(X_1^{k-1}, M_1^{k-1})$, we say that  $\mathcal{S}=\big(\cX, \{\mu_{k},d_{k}\}_{k=1}^{\infty},T\big)$ is a {\em non-adaptive} SQHT. Intuitively, a non-adaptive strategy is one used by the experimenter to choose the POVM $M_{k}$ at time $k$ {\em without} any  dependence on the past measurements and outcomes.

Let $\Omega :=(\M_{\mathcal{X}}\times \mathcal{X})^{\infty}$ be the infinite product space induced by  $\M_\mathcal{X}\times\mathcal{X}$ with the usual product topology. Let $\mathcal{F}$ be the $\sigma$-algebra generated by the product topology on $\Omega$. Given the underlying state $\rho$ and a sequence of adaptive strategies (as described above), we can define a probability measure $\P_{\mathcal{S},\rho}$ on $(\Omega,\mathcal{F})$ as follows. For any $k \in \mathbb{N}$, any measurable set $A\in\M_{\mathcal{X}}^k$ and any sequence~$x_1^k$, 
\begin{align}\label{Pmeasure}
\P_{\mathcal{S}, \rho} \big[(A\times x_1^k) \times (\M_{\mathcal{X}}\times \mathcal{X})^{\infty}\big]=\int_{A}\prod_{j=1}^{k}\mu_j(\mathrm{d}m_j | x_1^{j-1},m_{1}^{j-1})\Tr\big[\rho m_{j}(x_j)\big].
\end{align}  
The existence of the probability measure $\P_{\mathcal{S},\rho}$ can be justified by   Kolmogorov's extension theorem~\cite[A3, pp.~471]{Durrettprobability}. Then we can define the random process $\{(M_{k},X_k)\}_{k=1}^{\infty}$ with $(M_{k},X_k)$ being the coordinate map from $\Omega$ to $\M_{\mathcal{X}}\times\mathcal{X}$. Let $\mathcal{F}_k\subset\mathcal{F}$ be the $\sigma$-algebra generated by $(M_1^k,X_1^k)$. The event $\{T=k\}=\{d_{1}=\ldots=d_{k-1}=*, d_k\not=*\}$ that the experimenter stops the test at time $k$ is determined by $(M_1^k, X_1^k)$. Hence, for each $k\in\mathbb{N}$, the event $\{T=k\}$ belongs to~$\mathcal{F}_k$. Therefore, $T$ is a {\em stopping time} with respect to the filtration $\{\mathcal{F}_k\}_{k=1}^{\infty}$ (for more details on the definition and properties of a stopping time we refer the reader to~\cite[pp.~220]{Durrettprobability}).

\subsection{Sequential Quantum Probability Ratio Tests}\label{subsec:sprt}
In this subsection, we introduce the notion of a {\em sequential quantum probability ratio test} (SQPRT). Without loss of generality and for notational convenience, we assume in this section that for $k\ge 1$, let $\mu_{k}$  be the conditional probability mass function according to which the experimenter chooses POVM $M_{k}$ at time $k$.  Note that the probability measure defined through~(\refeq{Pmeasure}) only depends on the testing strategies $\{\mu_{k}\}_{k=1}^{\infty}$ and the underlying quantum state $\rho$. In the following when the testing strategies $\{\mu_{k}\}_{k=1}^{\infty}$ are given and the underlying state is $\rho_{i}$, we denote the probability measure $\P_{\mathcal{S},\rho_{i}}$ defined through~(\refeq{Pmeasure}) as $\P_{i}$ to simplify notation.  Let
\begin{align}\label{general:sprt}
S_{k}:=\log\frac{\P_{0}(X_{1}^{k},M_{1}^{k})}{\P_{1}(X_{1}^{k},M_{1}^{k})}.
\end{align}
Additionally, let $A$ and $B$ be two fixed positive real numbers. The decision function $d_{k}$ at time $k$ is defined as follows
 \begin{align}\label{eqn:general:sprt}
d_{k}(X_{1}^{k},M_{1}^{k})=\begin{cases}
0&S_{k}\ge B\\
1&S_{k}\le -A\\
*&\mbox{otherwise}.
\end{cases}
\end{align}
Let $T=\inf\{k\ge 1:S_{k}\not\in(-A,B)\}$ be the first time $k$ that $d_{k}\not=*$.  
Thus, $T$ is a stopping time with respect to the filtration generated by $\{ (M_{k}, X_k) \}_{k=1}^\infty$ as for any positive integer $k$, the event $\{T=k\}$ depends only on $(M_{1}^{k},X_{1}^{k})$, the first $k$ POVMs and outcomes. 

Intuitively, the experimenter keeps asking the Oracle for a new quantum state until $S_k$ is either larger than $B$ or smaller than $-A$. At this point in time, the experimenter is confident in making a definitive decision. The experimenter decides that  $\rho_0$ is the underlying state when $S_{T}\ge B$; otherwise, the experimenter decides that  $\rho_1$ is the underlying state. We say $\big(\cX,\{\mu_{k},d_{k}\}_{k=1}^{\infty},T\big)$ with $d_k$ specified as in~\eqref{eqn:general:sprt} is a {\em sequential quantum probability ratio test} with parameters $A$ and $B$. 
\subsection{Constraints and Achievable Exponents}

In the following we study sequences of SQHT $\mathcal{S}_n$, indexed by $n \in \mathbb{N}$. To simplify notation we use
$\P_{n,i}$ to denote $\P_{\mathcal{S}_n, \rho_i}$ for $i \in \{0,1\}$. The notation $\E_{n,i}[\cdot]$ means that the expectation is taken with respect to the probability measure $\P_{n,i}$.
We consider two types of constraints on the number of states $T_{n}$ used during the test. The first type of constraint is the {\em expectation constraint}: 
\begin{align}\label{expconstraint}
\max_{i \in \{0,1\}}\E_{n,i}[T_{n}]\le n.
\end{align} In other words, the average number of copies used in the testing procedure should be bounded by some  number $n$. The second type of constraint is the {\em probabilistic constraint}~\cite{anusha,litan}
\begin{align}\label{proconstraint}
\max_{i \in \{0,1\}}\P_{n,i}(T_{n}>n)<\eps
\end{align} for some fixed $\eps \in (0,1)$. In other words, the number of copies of the state used during the testing procedure should be bounded by some  number $n$ with probability larger than $1-\eps$. 

We study the trade-off between the error probabilities $(\alpha_n,\beta_n)$ under either the expectation or the probabilistic constraint on the number of copies of the state used during the test procedure. The first type of error is quantified by the probability that the experimenter declares that hypothesis $1$ is in effect when, in fact, hypothesis $0$ is true, i.e., 
\begin{equation}
\alpha_n := \P_{n,0}(d_{T_{n}} = 1). \label{eqn:alpha}
\end{equation}
On the other hand, the second type of error probability is 
\begin{equation}
\beta_n := \P_{n,1}(d_{T_{n}} = 0). \label{eqn:beta}
\end{equation}

\begin{definition}[Achievable Error Exponent Pairs] \label{def:ach_ee}
A pair $(R_0, R_1) \in\mathbb{R}_+^2$ is said to be {\em an achievable error exponent pair under the expectation constraint} if there exists a sequence of SQHTs $\{\mathcal{S}_n\}_{n\in\mathbb{N}}$ such that  
\begin{align}
\liminf_{n\to\infty}\frac{1}{n}\log\frac{1}{\alpha_n} & \ge R_0, \label{eqn:R0} \\ 
\liminf_{n\to\infty}\frac{1}{n}\log\frac{1}{\beta_n} &\ge  R_1, \quad \mbox{and}\label{eqn:R1} \\
\limsup_{n\to\infty}\Big( \max_{i \in \{0,1\}} \E_{n,i} [T_{n}]  -n\Big)&\le 0.  \label{eqn:exp_con}
\end{align}

Similarly, for $0<\eps<1$, a  pair $(R_0, R_1) \in \mathbb{R}_+^2$ is said to be {\em an $\eps$-achievable error exponent pair under the probabilistic constraint}  if there exists a  sequence of SQHTs $\{\mathcal{S}_n\}_{n\in\mathbb{N}}$ such that~\eqref{eqn:R0} and~\eqref{eqn:R1} hold and (instead of~\eqref{eqn:exp_con}),
\begin{equation}
\limsup_{n\to\infty} \max_{i \in \{0,1\}} \P_{n,i}(T_{n}>n)< \eps. \label{eqn:prob_con}
\end{equation}
\end{definition}
The condition in \eqref{eqn:exp_con} states that regardless of which hypothesis $i\in  \{0,1\}$ is in effect, the expectation value of the stopping time $\E_{n,i}[T_n]$ should not exceed $n+\gamma$ for any $\gamma>0$ for all $n$ sufficiently large.  In other words, we are allowing some additive slack on $\E_{n,i}[T_n]$.
\begin{definition}[Error Exponent Regions]
Define $\mathcal{A}_\mathrm{E} (\rho_0,\rho_1)\subset\mathbb{R}_+^2$, the {\em error exponent region under the expectation constraint}, to be the closure of the  set of all achievable error exponent pairs under the expectation constraint. 

Similarly, define $\mathcal{A}_\mathrm{P} (\eps|\rho_0,\rho_1)\subset\mathbb{R}_+^2$,  the {\em error exponent region under the $\eps$-probabilistic constraint}, to be the closure of the  set of all $\eps$-achievable error exponent pairs under the probabilistic  constraint.

Define $\mathcal{R}_\mathrm{E}(\rho_{0},\rho_{1})\subset \mathbb{R}^{2}_{+}$ and $\mathcal{R}_\mathrm{P}(\eps|\rho_0, \rho_1) \subset\mathbb{R}_+^2$ to be the sets of achievable error exponent pairs using {\em non-adaptive strategies} under the expectation and probabilistic constraints in~\eqref{eqn:exp_con} and~\eqref{eqn:prob_con}, respectively.  
\end{definition}

In the sequel, since $\rho_0 $ and $\rho_1$ are fixed, the explicit dependence on the states is often dropped from the notation for the error exponent regions.

\subsection{Information Quantities}

Consider two quantum states $\rho_0$ and $\rho_1$ with full support. Our results are stated in terms of the \emph{quantum relative entropy},
\begin{align}
	D(\rho_0\| \rho_1) := \Tr\left[ \rho_0 \left( \log \rho_0 - \log \rho_1 \right) \right] \,.
\end{align}
This is a generalization of the classical Kulback-Leibler divergence, which is recovered when $\rho_0$ and $\rho_1$ commute. Another generalization of the latter quantity is given by the {\em measured relative entropy}, which is defined as
\begin{align}\label{def:mre}
	D_{\M}(\rho_0\|\rho_1) := \sup_{m} D(P_{\rho_0,m}\|P_{\rho_1, m}),
\end{align}
where the supremum runs over all rank-1 PVMs comprised of $d$ projectors. The data-processing inequality for the quantum relative entropy ensures that $D_{\M}(\rho_0\|\rho_1)  \leq D(\rho_0\| \rho_1)$. Moreover, by~\cite[Theorem 2]{Marco2017}, we have
\begin{align}
	D_{\M}(\rho_0\|\rho_1) \geq D(P_{\rho_0,m}\|P_{\rho_1, m})
\end{align}
for any finite set $\mathcal{X}$ and any POVM $m \in \mathcal{M}_{\mathcal{X}}$. This means that the optimization in the definition of $D_{\M}$ can be extended to all POVMs without changing its value, which is key to the proof of the converse of Theorem~\ref{expectation}. Therefore we restate~\cite[Theorem 2]{Marco2017} as follows.

\begin{thm}\label{theorem:mre}
For two states $\rho_{0}$ and $\rho_{1}$ with full support, we have
\begin{align}
D_{\M}(\rho_0\|\rho_1)=\sup_{\mathcal{X}}\sup_{m\in\M_{\mathcal{X}}} D(P_{\rho_0,m}\|P_{\rho_1, m})
\end{align}
and the supremum is achieved at some PVM $m^{*}$ with $|\mathcal{X}|=d$.
\end{thm}

\section{Main Results}\label{main}

\subsection{Error Exponent Regions with Adaptive Testing Strategies} \label{sec:adapt}

We first state our main result for sequential quantum  hypothesis testing under the expectation constraints.
\begin{thm}\label{expectation}
Let $\rho_0$ and $\rho_1$ be two quantum states with full support. Then 
\begin{align}
\mathcal{A}_\mathrm{E}=\left\{(R_0,R_1):\begin{array}{c}
R_0\le D_{\M}(\rho_1\|\rho_0)\\
R_1\le D_\M(\rho_0\|\rho_1)
\end{array}\right\}.
\end{align}
\end{thm}
Our second results is an explicit characterization of $\mathcal{A}_{P}(\eps)$.
\begin{thm}\label{probabilistic}
Let $\rho_0$ and $\rho_1$ be two quantum states with full support. Then for any $0<\eps<1$, 
\begin{align}
\mathcal{A}_\mathrm{P}(\eps)=\left\{(R_0,R_1):\begin{array}{c}
R_0\le D_{\M}(\rho_1\|\rho_0)\\
R_1\le D_\M(\rho_0\|\rho_1)
\end{array}\right\}.
\end{align}
\end{thm}
All proofs of the theorems are deferred to Section~\ref{sec:proof}. Useful preparatory results for the proofs are collated in Section~\ref{sec:preli}.

In Theorems~\ref{expectation} and~\ref{probabilistic}, we derive the maximal achievable regions of the error exponents for separable adaptive measurements for the  sequential binary quantum hypothesis testing problem.  These imply that  $\mathcal{A}_\mathrm{E}$ and $\mathcal{A}_\mathrm{P}(\eps)$ are identical and characterized by a rectangle  whose top-right corner is given by the pair of measured relative entropies. Furthermore, we have also shown in Theorem \ref{probabilistic} what is known in information theory parlance as the {\em strong converse}. Namely, $\mathcal{A}_\mathrm{P}(\eps)$ does not depend on the permissible error probability $\eps\in(0,1)$. 

Since in practice there is no reason for the experimental to work on one state at a time, we consider the ``block'' version of the binary hypothesis test to examine the potential gains this framework yields. Consider the binary quantum hypothesis test, 
\begin{align}
H_0^{(l)}: \rho^{\otimes l}=\rho_0^{\otimes l}\qquad H_1^{(l)}: \rho^{\otimes l}=\rho_1^{\otimes l}.
\end{align} Under this setup, instead of requesting a single copy of the underlying state $\rho$, the experimenter   requests the Oracle to prepare and present $l$ samples of the underlying state $\rho^{\otimes l}$ at each point in time. As was done in Section~\ref{sec:model}, we can define the achievable regions of the error exponent pairs $\mathcal{A}_\mathrm{E}^{(l)}$ and $\mathcal{A}_\mathrm{P}^{(l)}(\eps)$ under the expectation and probabilistic constraints, respectively. Similar to Theorems~\ref{expectation} and~\ref{probabilistic}, we have  
\begin{align}
\mathcal{A}_\mathrm{E}^{(l)}=\mathcal{A}_\mathrm{P}^{(l)}(\eps)=\left\{(R_0,R_1):\begin{array}{c}
R_0\le \displaystyle\frac{1}{l} {D_{\M}(\rho_1^{\otimes l}\|\rho_0^{\otimes l})}\vspace{.5em}\\
R_1\le \displaystyle\frac{1}{l} {D_{\M}(\rho_0^{\otimes l}\|\rho_1^{\otimes l})}
\end{array}\right\}.
\end{align}

From~\cite{Petz1991}, it follows that
\begin{align}
\lim_{l\to\infty} \frac{D_{\M}(\rho_1^{\otimes l}\|\rho_0^{\otimes l})}{l}=D(\rho_1\|\rho_0)\quad\mbox{and}\quad \lim_{l\to\infty} \frac{D_{\M}(\rho_0^{\otimes l}\|\rho_1^{\otimes l})}{l}=D(\rho_0\|\rho_1).
\end{align}
Using this limiting relation, we can characterize the ultimate quantum limit of achievable error exponent pairs using sequential adaptive testing strategies as follows. 
\begin{thm}\label{ultimate:adaptive}
Let $\rho_0$ and $\rho_1$ be two quantum states with full support. Then for any $0<\eps<1$, 
\begin{align}
\bigcup_{l=1}^{\infty}\mathcal{A}_\mathrm{E}^{(l)}=\bigcup_{l=1}^{\infty}\mathcal{A}_\mathrm{P}^{(l)}(\eps)=\left\{(R_0,R_1):\begin{array}{c}
R_0 \le D(\rho_1\|\rho_0)\\
\, R_1 \le D(\rho_0\|\rho_1)
\end{array}\right\}.
\end{align}
\end{thm}
\subsection{Error Exponent Regions with Non-Adaptive Testing Strategies} \label{sec:non-adapt}
In this section we state our results for $\mathcal{R}_{\mathrm{E}}$ and $\mathcal{R}_{\mathrm{P}}(\eps)$, the regions of error exponent pairs when {\em non-adaptive} tests are permitted. For any subset $A$ of the plane $\mathbb{R}^{2}$, let $\overline{\mathrm{Conv}(A)}$ be the closure of the convex hull of $A$.  The following two theorems fully characterize   $\mathcal{R}_\mathrm{E}$ and $\mathcal{R}_\mathrm{P}(\eps)$.

\begin{thm}\label{non-adaptive}
Let $\rho_0$ and $\rho_1$ be two quantum states with full support. Then for any $0<\eps<1$, 
\begin{align}
 \mathcal{R}_\mathrm{E}=\mathcal{R}_{\mathrm{P}}(\eps)=\overline{\mathrm{Conv}(\mathcal{C})},
\end{align}
where 
\begin{align}\mathcal{C}=\bigcup_{\mathcal{X}}\bigcup_{m\in\M_\mathcal{X}}\left\{(R_0,R_1):\begin{array}{c}
R_0\le D(P_{\rho_1,m}\|P_{\rho_0, m})\\
R_1\le D(P_{\rho_0,m}\|P_{\rho_1, m})
\end{array}
\right\},\end{align}
and $\mathcal{X}$ runs over all finite sets and $\mathcal{M}_{\mathcal{X}}$ is the set of POVMs with support $\mathcal{X}$.
\end{thm} 
Similar to the adaptive case, we may apply non-adaptive strategies to $l$ samples of the given state. We    define the regions of achievable error exponent pairs $\mathcal{R}_\mathrm{E}^{(l)}$ and $\mathcal{R}_{\mathrm{P}}^{(l)}(\eps)$ under the expectation and probabilistic constraints, respectively. Additionally, let $\mathcal{M}_\mathcal{X}^{(l)}$ be the set of POVMs indexed by $\mathcal{X}$ performed on the system $\mathcal{H}^{\otimes l}$. We have the following characterization of the ultimate quantum limit using non-adaptive strategies for  SQHTs.
\begin{corollary}
Let $\rho_0$ and $\rho_1$ be two quantum states with full support and let
\begin{align}
\mathcal{C}^{(l)}=\bigcup_{\mathcal{X}}\bigcup_{m\in\M_\mathcal{X}^{(l)}}\left\{(R_0,R_1):\begin{array}{c}
\displaystyle R_0\le \frac{1}{l} D\big(P_{\rho_1^{\otimes l},m}\|P_{\rho_{0}^{\otimes l}, m}\big) \vspace{.5em}\\
\displaystyle R_1\le \frac{1}{l} D\big(P_{\rho_0^{\otimes l},m}\|P_{\rho_{1}^{\otimes l}, m}\big)
\end{array}
\right\}.
\end{align}
Then for any $0<\eps<1$, we have
\begin{align}
\mathcal{R}_\mathrm{E}^{(l)}=\mathcal{R}_{\mathrm{P}}^{(l)}(\eps)=\overline{\mathrm{Conv}(\mathcal{C}^{(l)})},
\end{align}
and
\begin{align}
\bigcup_{l=1}^{\infty}\mathcal{R}_\mathrm{E}^{(l)}=\bigcup_{l=1}^{\infty}\mathcal{R}_{\mathrm{P}}^{(l)}(\eps).
\end{align}
\end{corollary}

\subsection{Discussions on the Main Results} \label{sec:discuss}
In Theorems~\ref{expectation} and \ref{probabilistic}, complete characterizations of the regions of achievable error exponent pairs under both types of constraints are provided. We contrast our results to those in \cite{quantumSHT} in this section. In particular, in~\cite[Equation~(11)]{quantumSHT}, the authors showed that for any fixed POVM $m$ and  using a sequence of  SPRT,   the expected number of samples needed to achieve vanishing error probabilities $\alpha$ and $\beta$  behave as 
\begin{align}
\E_0[T]= \frac{1+o(1)}{D(P_{\rho_0,m}\|P_{\rho_1,m})}\log\frac{1}{\beta},\\ 
\E_1[T]= \frac{1+o(1)}{D(P_{\rho_1,m}\|P_{\rho_0,m})}\log\frac{1}{\alpha}.
\end{align}
However, if one uses  adaptive protocols and a sequence of  SQPRTs in the proof of Theorem~\ref{expectation}, for vanishing error probabilities $\alpha$ and $\beta$, we have that
\begin{align}
\E_0[T]= \frac{1+o(1)}{D_{\M}(\rho_1\|\rho_0)}\log\frac{1}{\beta},\\
\E_1[T]= \frac{1+o(1)}{D_{\M}(\rho_0\|\rho_1)}\log\frac{1}{ \alpha}.
\end{align}
We note that our result  strictly improves on that in \cite{quantumSHT} apart from the scenario in which there exists a POVM $m$ that {\em simultaneously} achieves the suprema in the definitions of  $D_{\M}(\rho_0\|\rho_1)$ and $D_{\M}(\rho_1\|\rho_0)$ in~\eqref{def:mre}. 

Using the ``block" POVM $m$ operating on $l$ independent samples of the state $\rho^{\otimes l}$ with $l\to \infty$, the authors in~\cite[Theorem 1]{quantumSHT} also showed that as $\max\{\alpha,\beta\}\to 0$,
\begin{align}
\E_0[T]\ge \frac{1+o(1)}{D(\rho_0\|\rho_1)}\log\frac{1}{\beta},\label{eqn:one}\\
\E_1[T]\ge \frac{1+o(1)}{D(\rho_1\|\rho_0)}\log\frac{1}{\alpha}.\label{eqn:two}
\end{align} 
However, the question of the existence of a  sequence of SQHTs that {\em simultaneously} achieves the lower bounds in (\refeq{eqn:one}) and (\refeq{eqn:two}) was left unanswered in~\cite{quantumSHT}. We answer this in the affirmative in Theorem~\ref{ultimate:adaptive}.

\subsection{Numerical Example}\label{sec:num}
Before we present the numerical results, we first show that to evaluate the region $\overline{\mathrm{Conv}(\mathcal{C})}$ in Theorem~\ref{non-adaptive} it suffices to consider the POVMs with at most $d^{2}$ elements. Let $[d^{2}]=\{1,2,\ldots,d^{2}\}$ and let $
\M_{[d^{2}]}^{(1)}=\{m\in \M_{[d^{2}]}: m(x)\ \mbox{is of rank one for all $x\in [d^{2}]$}\}$.
\begin{thm}\label{thm:card}
Let $\rho_0$ and $\rho_1$ be two quantum states with full support. Then 
\begin{align}
\overline{\mathrm{Conv}(\mathcal{C})}=\overline{\mathrm{Conv}(\mathcal{C}^{(1)})}\end{align}
where 
\begin{align}
\mathcal{C}^{(1)}=\bigcup_{m\in\M_{[d^{2}]}^{(1)}}\left\{(R_0,R_1):\begin{array}{c}
R_0\le D(P_{\rho_1,m}\|P_{\rho_0, m})\\
R_1\le D(P_{\rho_0,m}\|P_{\rho_1, m})
\end{array}
\right\}.
\end{align}
\end{thm}
The proof of Theorem~\ref{thm:card} can be found in Subsection~\ref{subsec:thm:card}.
Now we provide an example  inspired by~\cite{quantumSHT} to illustrate the advantage of adaptive strategies over non-adaptive ones.

For this purpose, let $\rho_0=r_0\ket{\psi_0}\bra{\psi_0}+(1-r_0)\frac{I}{2}$ and $\rho_1=r_1\ket{\psi_1}\bra{\psi_1}+(1-r_1)\frac{\id}{2}$, where $\ket{\psi_i}=\cos\frac{\theta}{4}\ket{0}+(-1)^{i}\sin\frac{\theta}{4}\ket{1}$, $0\le \theta\le \pi$, and $0\le r_i\le 1$, $\ket{0}=(1,0)^\top$, $\ket{1}=(0,1)^\top$, $I$ is the $2\times 2$ identity matrix. For $\rho_0$ and $\rho_1$ with parameters $(r_0,r_1,\theta)$, we define the sum rate of error exponent pairs as follows: 
\begin{align}
	f(r_0,r_1,\theta):=D_\M(\rho_1\|\rho_0)+D_\M(\rho_0\|\rho_1)
\end{align} 
and 
\begin{align}\label{eqn:hatg}
g(r_0,r_1,\theta):=\sup_{\mathcal{X}}\sup_{m\in\M_{\mathcal{X}}}D(P_{\rho_0,m}\|P_{\rho_1,m})+D(P_{\rho_1,m}\|P_{\rho_0,m}).
\end{align}
From Theorem~\ref{thm:card} it follows that
\begin{align}\label{eqn:hatg2}
g(r_0,r_1,\theta)=\sup_{m\in\M_{[d^{2}]}^{(1)}}D(P_{\rho_0,m}\|P_{\rho_1,m})+D(P_{\rho_1,m}\|P_{\rho_0,m}).
\end{align}

\begin{figure}[t]
\centering
   \begin{minipage}{0.48\textwidth}
     \vspace{0.2cm}
     \centering
    \includegraphics[width =1\columnwidth]{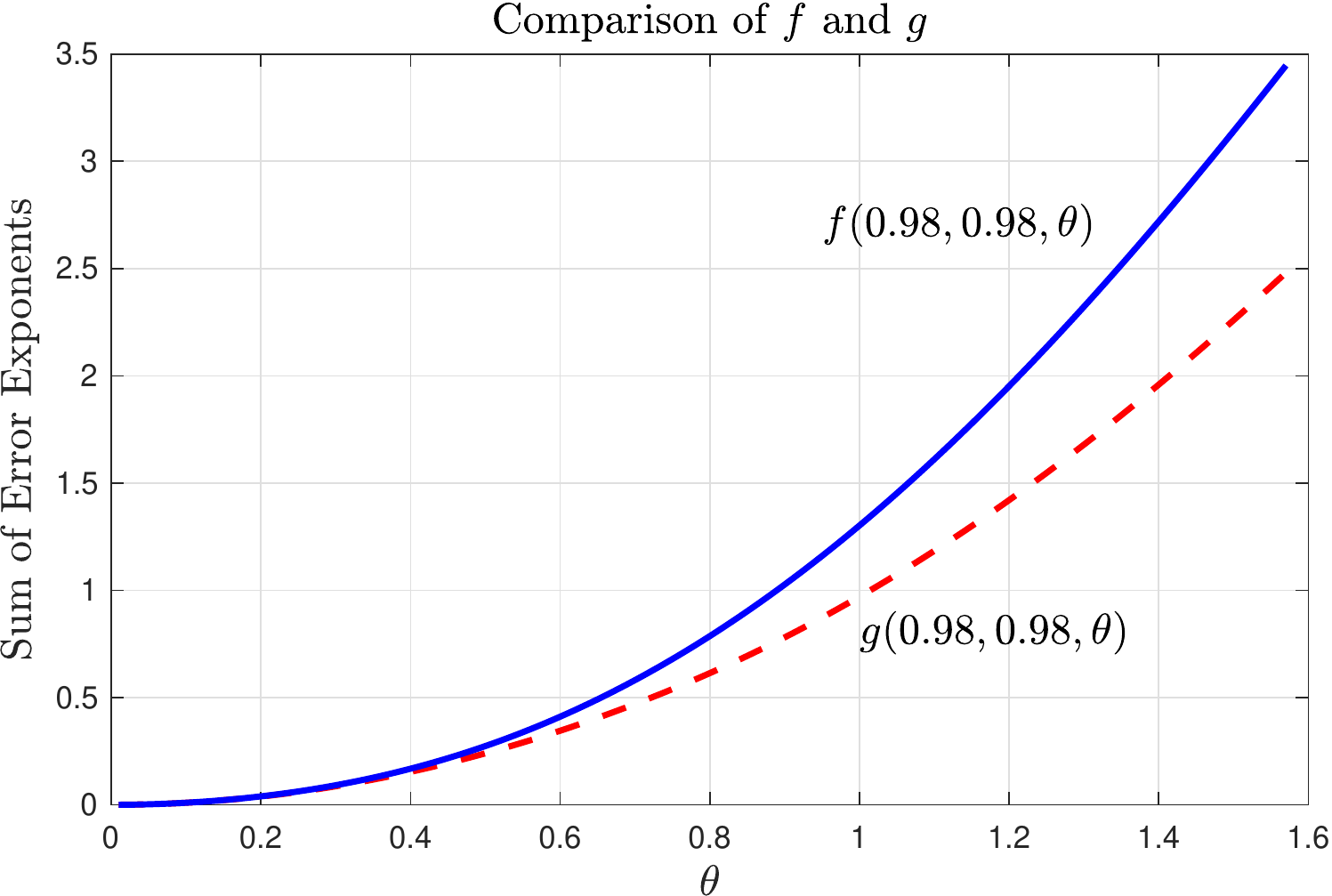}
 \caption{Maxima of the sum rate of error exponent pairs with adaptive or non-adaptive measurement strategies for $r_1=r_2=0.98$ and $\theta\in (0,\frac{\pi}{2})$. The gap is most pronounced for large values of $\theta$.}
 \label{fig:sumrate}
   \end{minipage}\hspace{5mm}
   \begin{minipage}{0.48\textwidth}
     \centering
\begin{overpic}[width =.85\columnwidth]{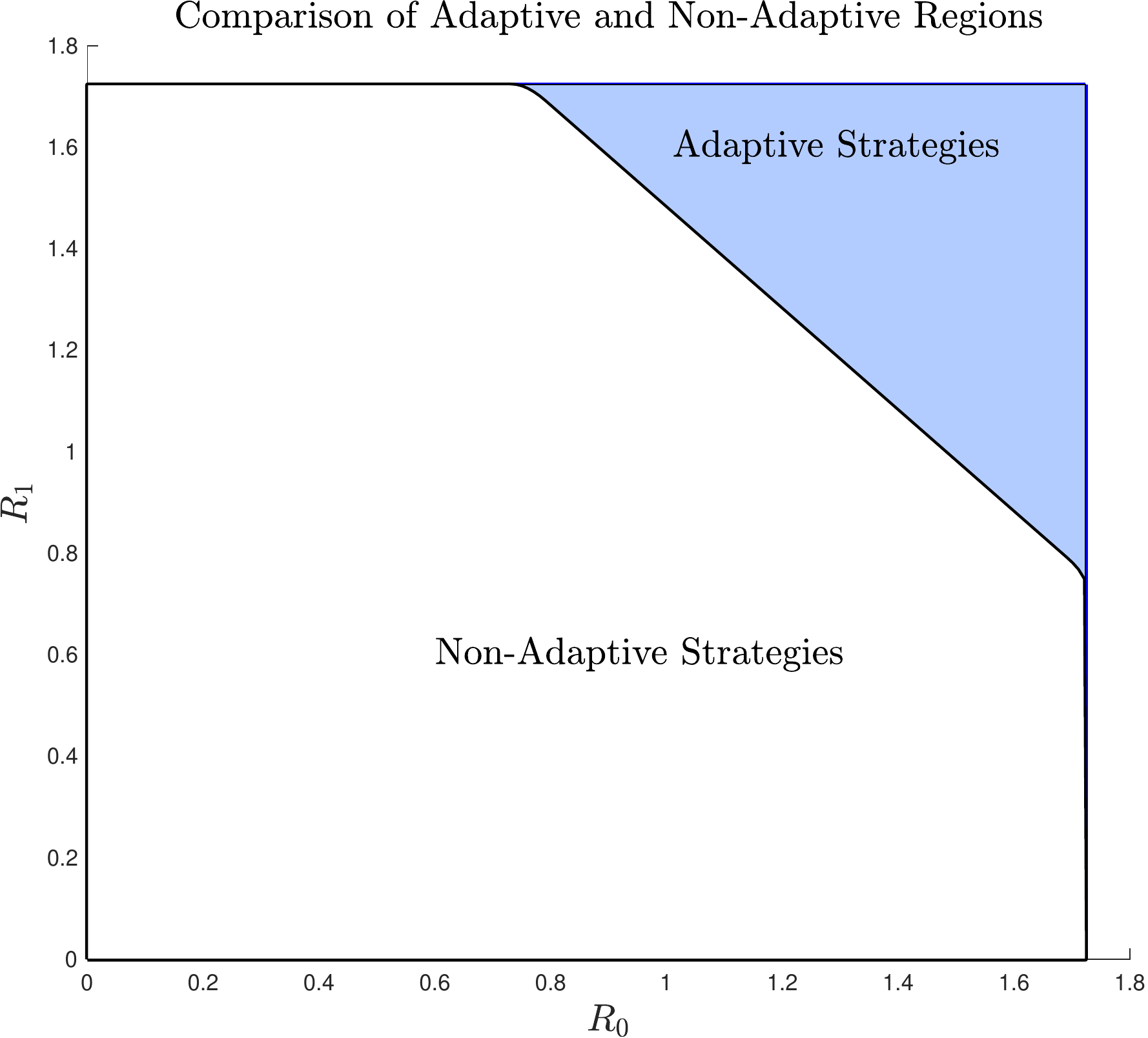}
 \put(96,10){{\footnotesize  $D_{\mathcal{M}}(\rho_1\|\rho_0)$}}
 \put(95,7){\circle*{2}}
 \put(8, 83){\circle*{2}}
 \put(-19,81){{\footnotesize  $D_{\mathcal{M}}(\rho_0\|\rho_1)$}}
   \end{overpic}
\caption{Achievable regions of error exponent pairs with adaptive or non-adaptive measurement strategies for when $(r_1,r_2, \theta)=(0.98,0.98,1.57)$. Note that the region for adaptive strategies is the entire rectangle including the region for non-adaptive strategies.}\label{fig:region}
   \end{minipage}
\hfill
\end{figure}

In Figure~\ref{fig:sumrate}, we numerically evaluate $f$ and $g$ for some parameter range. We observe a gap between the two quantities, which indicates that there is no single measurement that can simultaneously achieve the measured relative entropies $D_\M(\rho_1\|\rho_0)$ and $D_\M(\rho_0\|\rho_1)$. Figure~\ref{fig:sumrate} thus shows that adaptive measurements yield smaller error probabilities (larger error exponents) vis-\`a-vis non-adaptive measurements. The maximal achievable regions of the error exponent pairs using adaptive and non-adaptive measurement strategies are then numerically evaluated in Figure~\ref{fig:region}. This figure corroborates the superiority of adaptive measurements over their non-adaptive counterparts.

\section{Preliminaries}\label{sec:preli}
In this section, we collect some known results used in the proof of the main results. For any set $A$, we use $\chi_A$ to denote the indicator function of the set $A$. We first recapitulate the definitions of {\em conditional expectation} and {\em submartingales}.
\begin{definition}
Let $(\Omega,\mathcal{F},\mu)$ be a probability space. Let $X$ be a random variable with $\E[|X|]<\infty$. Let $\mathcal{G}\subset\mathcal{F}$ be a sub-$\sigma$-algebra. The {\em conditional expectation} $\E[X|\mathcal{G}]$ of $X$ given $\mathcal{G}$ is defined as a $\mathcal{G}$-measurable random variable such that
\begin{align}\label{conditional}
\E[X\chi_{A}]=\E\big[\E[X|\mathcal{G}]\chi_A\big]\quad\mbox{for any $A\in\mathcal{G}$.}
\end{align}
\end{definition}
\begin{remark}
There are numerous $\mathcal{G}$-measurable random variables that satisfy~(\refeq{conditional}); they are called {\em versions} of the conditional expectation of $X$ with respect to $\mathcal{G}$. However, any two of them are equal almost surely. In this sense, the conditional expectation is uniquely defined through (\refeq{conditional}).
\end{remark}
\begin{definition}
A discrete-time stochastic process $\{X_k\}_{k=1}^{\infty}$ is called a {\em submartingale} if $\E[|X_k|]<\infty$ and $\E[X_k|\mathcal{F}(X_1^{k-1})]\ge X_{k-1}$ almost surely. Similarly,     $\{X_k\}_{k=1}^{\infty}$ is called a {\em supermartingale} if $\{-X_k\}_{k=1}^{\infty}$ is  a submartingale. Finally, $\{X_k\}_{k=1}^{\infty}$ is called a {\em martingale} if %it is both a submartingale and a supermatrigale.  
$\{X_k\}_{k=1}^{\infty}$ and $\{-X_k\}_{k=1}^{\infty}$ are submartingales.
\end{definition}

\begin{thm}\cite[Theorem 4.8.5, pp.~256]{Durrettprobability}\label{optionalstopping}
Suppose that $\{X_k\}_{k=1}^{\infty}$ is a submartingale (resp.\ supermartingale) and $\E\big[|X_{k+1}-X_k|\big|\mathcal{F}(X_1^{k})\big]\le C$ almost surely for some finite constant $C$. If $T$ is a stopping time with $\E[T]<\infty$, then $\E[X_T]\ge \E[X_1]$ (resp.\ $\E[X_T]\le \E[X_1]$).
\end{thm}
The following theorem known as Doob's maximal inequality bounds the tail probabilities of the maximum of a collection of non-negative submartingales.
\begin{thm}\label{doob}\cite[Theorem 4.4.2, pp.~235]{Durrettprobability}
Suppose $\{X_k\}_{k=1}^{\infty}$ is a non-negative submartingale. Then for any $\lambda>0$, 
\begin{align}
     \Pr\left(\max_{1\le j\le k}X_j\ge \lambda\right)\le \frac{\E[X_k]}{\lambda}.
\end{align}
\end{thm}

The following lemma provides an upper bound on the logarithm of the likelihood ratio by the max-relative entropy. 
\begin{lemma}\label{lemma:mre}
Let $\rho_0$ and $\rho_1$ be two quantum states such that $\mathrm{Supp}(\rho)=\mathrm{Supp}(\sigma)=\mathcal{H}$. Then
for any non-zero positive semidefinite matrix ${\bf A}$, we have 
\begin{equation}
\big|\log{\Tr[{\bf A}\rho_0]}-\log{\Tr[{\bf A}\rho_1]}\big|\le \max\big\{D_{\mathrm{max}}(\rho_0\|\rho_1),D_{\mathrm{max}}(\rho_1\|\rho_0) \big\} =: C,
\end{equation}
where \begin{align}
D_{\mathrm{max}}(\rho\|\sigma):=\log\inf\{\lambda>0: \rho\le \lambda \sigma\}.
\end{align}
is the max-relative entropy between $\rho_{0}$ and $\rho_{1}$.

\end{lemma}

\begin{proof}
The statement follows directly from the definition of the max-relative entropy.
\end{proof}

Assume $\rho_{0}$ and $\rho_{1}$ have full support. Let $\mathcal{S}=\big(\cX, \{\mu_{k},d_{k}\}_{k=1}^{\infty},T\big)$ be an SQHT as defined in Subsection~\ref{subsec:sqht}. For $i\in\{0, 1\}$ and $k\ge 1$, let $p_{i,k}(\cdot\mid\cdot)$ be the conditional probability measure of $(M_{k},X_k)$ given $\{(M_{j},X_j)\}_{j=1}^{k-1}$ when the underlying state is $\rho_i$. As the conditional probability of $X_{k}=x$ given $(X_1^{k-1},M_{1}^{k})$ is $\Tr\big[\rho_i M_{k}(x)\big]$, we have that, for any $M_{k}=m_{k}$ and $X_{k}=x_{k}$,
\begin{align}
p_{i,k}(\mathrm{d}m_{k},x_k|X_1^{k-1},M_{1}^{k-1})&=\mu_{k}(\mathrm{d}m_{k}|X_1^{k-1},M_{1}^{k-1})\Tr \big[\rho_i m_{k}(x_k) \big].
\end{align}
Let 
\begin{align}\label{eqn:llr}
Z_k:=\log{\Tr \big[\rho_0 M_{k}(X_k) \big]}-\log{\Tr \big[\rho_1 M_{k}(X_k) \big]}.
\end{align} 
Note that the conditional expectation of $Z_k$ given $\{M_k=m\}$ satisfies
\begin{align}
\E_{0}[Z_{k}|M_{k}=m]=D(P_{\rho_{0},m}\|P_{\rho_{1},m}).\label{eqn:condexp}
\end{align}
We also observe that $(X_1^{k-1}, M_1^{k-1}) - M_{k}- Z_k$ forms a Markov chain. Recall from~(\refeq{general:sprt}) that
 $S_k$ is the logarithm of the likelihood ratio. Then using the chain rule for probability measures, we have that
\begin{align}
S_k&=\sum_{j=1}^{k}\log\frac{p_{0,j}(M_{j},X_j|X_1^{j-1},M_{1}^{j-1})}{p_{1,j}(M_{j},X_j|X_1^{j-1},M_{1}^{j-1})}=\sum_{j=1}^k Z_j,\label{eqn:sk}
\end{align}
where~(\refeq{eqn:sk}) follows from the definition of $Z_{j}$ in~(\refeq{eqn:llr}).

\begin{lemma}\label{lemma:summartingale}
Let $\rho_{0}$ and $\rho_{1}$ be two quantum states with full support. Then under hypothesis that $\rho=\rho_{0}$,
\begin{itemize}
\item [(i)]  The stochastic process $\{S_k-kD_{\M}(\rho_{0}\|\rho_1)\}_{k=1}^{\infty}$ is a supermartingale.
\item[(ii)]  The stochastic process $\{S_k\}_{k=1}^{\infty}$ is a submartingale.
\end{itemize}
\end{lemma}
\begin{proof}
We first prove Part (i).  Note that
\begin{align}
\E_0[S_k-&kD_{\M}(\rho_{0}\|\rho_1)|\mathcal{F}_{k-1}]-\big(S_{k-1}-(k-1)D_{\M}(\rho_{0}\|\rho_1)\big)\notag\\
&=\E_0[Z_k-D_{\M}(\rho_{0}\|\rho_1)|\mathcal{F}_{k-1}]\\
&=\int \mu_{k}(\mathrm{d}m_{k}|X_1^{k-1},M_1^{k-1}) \E_0[Z_k-D_{\M}(\rho_{0}\|\rho_1)|M_{k}=m_{k}]\label{eqn:super22}\\
&=\int \mu_{k}(\mathrm{d}m_{k}|X_1^{k-1},M_1^{k-1}) \big(D\big(P_{\rho_{0},m_{k}}\big\|P_{\rho_{1},m_{k}}\big)-D_{\M}(\rho_{0}\|\rho_1)\big)\\
&\le 0,\label{eqn:super23}
\end{align}
where~(\refeq{eqn:super22}) follows from the fact that $(X_1^{k-1},M_{1}^{k-1})-M_{k}-Z_k$ is a Markov chain and (\refeq{eqn:super23}) follows from Theorem~\ref{theorem:mre} and~\eqref{eqn:condexp}. Therefore $\{S_k-kD_{\M}(\rho_{0}\|\rho_1)\}_{k=1}^{\infty}$ is a supermartingale.

Now we prove Part (ii). Note that
\begin{align}
\E_0[S_k|\mathcal{F}_{k-1}]-S_{k-1}&=\E_0[Z_{k}|\mathcal{F}_{k-1}]\\
&=\int \mu_{k}(\mathrm{d}m_{k}|X_1^{k-1},M_1^{k-1}) \E_0[Z_k|M_{k}=m_{k}]\label{eqn:sub22}\\
&=\int \mu_{k}(\mathrm{d}m_{k}|X_1^{k-1},M_1^{k-1}) D\big(P_{\rho_{0},m_{k}}\big\|P_{\rho_1,m_{k}}\big)\label{eqn:sub23}\\
&\ge 0,
\end{align}
where~(\refeq{eqn:sub22}) follows from the fact that $(X_1^{k-1},M_{1}^{k-1})-M_{k}-Z_k$ is a Markov chain and (\refeq{eqn:sub23}) follows from \eqref{eqn:condexp}. Hence $\{S_k\}_{k=1}^{\infty}$ is a submartingale. 

This completes the proof of Part (ii) of Lemma~\ref{lemma:summartingale}. 
\end{proof}

The following lemma is used to derive bounds on the error probabilities in classical sequential hypothesis testing problems.

\begin{lemma}\label{lemma:general}
Let $\mu_{0}$ and $\mu_{1}$ be two probability measures over $(\Omega,\mathcal{F})$ and let $\{\mathcal{F}_{k}\}_{k=1}^{\infty}$ be a filtration. Let $\mu_{0}$ and $\mu_{1}$ be mutually absolutely continuous over $\mathcal{F}_{k}$ and let $G_{k}$ be the logarithm of the Radon-Nikodym derivative of $\mu_{0}$ with respect to $\mu_{1}$ over $\mathcal{F}_{k}$.  Let $T$ be a stopping time with respect to the filtration $\{\mathcal{F}_{k}\}_{k=1}^{\infty}$ and let $\mathcal{F}_{T}$ be the $\sigma$-algebra generated by $T$. Let $\delta$ be a $\{0,1\}$-valued $\mathcal{F}_{T}$ measurable function. Let $\alpha=\mu_{0}(\delta=1)$ and $\beta=\mu_{1}(\delta=0)$. 
Suppose $\min_{i=0,1}\mu_{i}(T<\infty)=1$. 
\begin{enumerate}[(i)]
\item\label{com} For any bounded $\mathcal{F}_T$-measurable random variable $Y$, we have
\begin{align}
    \E_{\mu_{0}}[Y]=\E_{\mu_{1}}[Ye^{-G_T}]\quad \mbox{and}\quad \E_{\mu_{1}}[Y]=\E_{\mu_{0}}[Ye^{-G_T}];
\end{align}
\item \label{lemma:probs_{k}}
For any $E\in \mathcal{F}_{T}$ and $\lambda>0$, the following inequalities hold,
\begin{align}
&\mu_{0}(E)-\lambda \mu_{1}(E)\le \mu_{0}(G_{T}\ge \log\lambda)\label{eqn:item:type1} \quad\mbox{and}\\
&\mu_{1}(E)-\lambda \mu_{0}(E)\le \mu_{1}(-G_{T}\ge \log\lambda).
\end{align}
\end{enumerate}
\end{lemma}
Part~(\refeq{com}) is from~\cite[Theorem 1.1, pp.~4]{nonlinearrenewaltheory}, Part~(\refeq{lemma:probs_{k}}) is from~\cite[Lemma 7]{litan}.
 We now apply Lemma~\ref{lemma:general} to an SQHT $\mathcal{S}$ to obtain the following corollary.
\begin{corollary}\label{cor:sht}
Assume $\rho_{0}$ and $\rho_{1}$ have full support. Let $\mathcal{S}=\big(\cX,\{\mu_{k},d_{k}\}_{k=1}^{\infty},T\big)$ be an SQHT  such that $\min_{i=0,1}\P_{i}(T<\infty)=1$.
\begin{enumerate}[(i)]
\item\label{com:sht} For any bounded $\mathcal{F}_T$-measurable random variable $Y$, we have
\begin{align}\label{eqn:item:com}
    \E_0[Y]=\E_1[Ye^{-S_T}]\quad \mbox{and}\quad \E_1[Y]=\E_0[Ye^{-S_T}].
\end{align}
\item \label{probs_{k}:sht}  For any $\lambda>0$, the following inequalities hold
\begin{align}
&\P_{0}(d_{T}=0)-\lambda \P_{1}(d_{T}=0)\le \P_{0}(S_{T}\ge \log\lambda)\label{item:eqn:type1} \quad\mbox{and}\\
&\P_{1}(d_{T}=1)-\lambda \P_{0}(d_{T}=1)\le \P_{1}(-S_{T}\ge \log\lambda)\label{item:eqn:type2}.
\end{align}
\end{enumerate}
\end{corollary} 
\begin{proof}
One can easily check that $\P_{0}$ and $\P_{1}$ defined in~(\refeq{Pmeasure}) are mutually absolutely continuous over $\mathcal{F}_{k}$ and that $S_{k}$ defined through~(\refeq{eqn:sk}) is the logarithm of the Radon-Nikodym derivative of $\P_{0}$ with respect to $\P_{1}$ over $\mathcal{F}_{k}$. As $
\min_{i=0,1}\P_{i}(T<\infty)=1$,  the conditions in Lemma~\ref{lemma:general} are satisfied. Therefore we obtain (\refeq{eqn:item:com})  by applying Part (i) of Lemma~\ref{lemma:general}. Letting $E=\{d_{T}=0\}$ in~(\refeq{eqn:item:type1}), we obtain~(\refeq{item:eqn:type1}). Analogously, we have~(\refeq{item:eqn:type2}).  
This completes the proof of Corollary~\ref{cor:sht}.
\end{proof} 
\section{Proofs of the Main Results}\label{sec:proof}
In the proof of achievability parts of the main results, we construct a sequence of SQPRTs $\{\mathcal{S}_{n}\}$ with $\mathcal{S}_{n}=(\cX, \{\mu_{k},d_{n,k}\}_{k=1}^{\infty},T_{n})$ that satisfies (\refeq{eqn:R0}),~(\refeq{eqn:R1}),~(\refeq{eqn:exp_con}),~and~(\refeq{eqn:prob_con}) for appropriate $R_{0}$ and $R_{1}$. Note that the strategies $\{\mu_{k}\}_{k=1}^{\infty}$ do  not depend on $n$. As $\P_{n,i}$ defined through~(\refeq{Pmeasure}) only depends on the strategies $\{\mu_{k}\}_{k=1}^{\infty}$ and the underlying state $\rho_{i}$, the probability measure $\P_{n,i}$ associated to each SQPRT $\mathcal{S}_{n}$ is the same. Therefore in the proof of achievability parts, we adopt the notations $\P_{i}$, $\E_{i}$ and $S_{k}$ instead of $\P_{n,i}$, $\E_{n,i}$ and $S_{n,k}$.

In the proof of the converse parts of the main results, we consider {\em arbitrary} sequences of SQHTs $\{\mathcal{S}_{n}\}_{n=1}^{\infty}$ with $\mathcal{S}_{n}=\big(\cX_n,\{\mu_{n,k},d_{n,k}\}_{k=1}^{\infty},T_{n}\big)$. Recall from Section~\ref{subsec:sqht} that $\P_{n,0}$ (resp.\ $\P_{n,1}$) is the probability measure defined by~(\refeq{Pmeasure}) on $\big(\mathcal{X}_{n}\times\M_{\mathcal{X}_{n}}\big)^{\infty}$ for the strategies $\{\mu_{n,k}\}_{k=1}^{\infty}$  when the underlying state is $\rho_{0}$ (resp.\ $\rho_{1}$). Let $\{(X_{n,k},M_{n,k})\}_{k=1}^{\infty}$ be the random process of the outcomes and measurements associated to the SQHT $\mathcal{S}_{n}$. Recall from~(\refeq{general:sprt}),~(\refeq{eqn:llr}) and~(\refeq{eqn:sk}) that 
\begin{align}\label{def:znj}
 Z_{n,j}&=\log \Tr\big[\rho_{0}M_{n,j}(X_{n,j})\big]-\log \Tr\big[\rho_{1}M_{n,j}(X_{n,j})\big]
 \end{align}
and
\begin{align}\label{def:snk}
  S_{n,k}&=\sum_{j=1}^{k}Z_{n,j},
\end{align}   
where $X_{n,1}^{k}=(X_{n,1},\ldots,X_{n,k})$.

\subsection{Achievability Proofs of Theorems~\ref{expectation} and~\ref{probabilistic}}

Before presenting the proof of Theorems~\ref{expectation} and~\ref{probabilistic}, we first introduce the sequence of SQPRTs that achieves $\big(D_\M(\rho_1\|\rho_0), D_\M(\rho_0\|\rho_1)\big)$.  Without loss of generality, let $\mathcal{X}=\{1,2,\ldots,d\}$. From the definition of the measured relative entropy, there exists two PVMs $m^{*}_0=\{m^{*}_{0}(x)\}_{x\in\mathcal{X}}$ and $m^*_1=\{m^*_{1}(x)\}_{x\in\mathcal{X}}$ that achieve the suprema in the definitions of $D_\M(\rho_0\|\rho_1)$ and $D_\M(\rho_1\|\rho_0)$, respectively. 

 Throughout this subsection, instead of $\mu_{k}$ we use $p_{k}$ to denote the adaptive strategies as the adaptive strategies we define in the following are probability mass functions. We now define the adaptive strategies $\{p_k\}_{k=1}^{\infty}$ used in the SQPRTs. For $k=1$ and $i\in\{0,1\}$, we set $p_1(m_i^{*})={1}/{2}$. That is, the experimenter at time $1$ chooses $M_{1}\in\{m_0^*,m_1^*\}$ uniformly at random. Recall that
$Z_j=\log{\Tr \big[\rho_0 M_{j}(X_j) \big]}-\log{\Tr \big[\rho_1 M_{j}(X_j) \big]}$ and $S_{k}=\sum_{j=1}^{k}Z_{j}$. 
For $k\ge 2$, the POVM $M_{k}$ is chosen by the experimenter at time $k$ according to the accumulated sum of log-likelihoods $S_{k}$ as follows
\begin{align}\label{dynamics}
M_{k}=\begin{cases}
m^*_0&\mbox{if}\ S_{k-1}\ge0\\
m^*_1&\mbox{otherwise}.
\end{cases}
\end{align}
Therefore for $k\ge 1$, the adaptive strategies are defined as follows
\begin{align}
p_{k}(m_0^{*}|x_{1}^{k-1},m_{1}^{k-1})&=\begin{cases}\frac{1}{2}&\mbox{if $k=1$}\\1&\mbox{if $k\ge 2$ and $S_{k-1}\ge 0$},\end{cases} \quad\mbox{and}\label{eqn:d1}\\
p_{k}(m_1^{*}|x_{1}^{k-1},m_{1}^{k-1})&=\begin{cases}\frac{1}{2}&\mbox{if $k=1$}\\1&\mbox{if $k\ge 2$ and $S_{k-1}< 0$}.\end{cases}\label{eqn:d2}
\end{align} 
 For any fixed $0<\tau<\min\{D_\M(\rho_1\|\rho_0), D_\M(\rho_0\|\rho_1)\}$, let 
 \begin{align}
 A_n:=n(D_\M(\rho_1\|\rho_0)-\tau)\quad\mbox{and}\quad B_n:=n(D_\M(\rho_0\|\rho_1)-\tau).
 \end{align} For any $n\ge 1$, let $\mathcal{S}_{n}=\big(  \cX,\{p_{k},d_{n,k}\}_{k=1}^{\infty},T_{n}\big)$ be the SQPRT with parameters $A_{n}$ and $B_{n}$. Recall from~(\refeq{eqn:general:sprt}) in Subsection~\ref{subsec:sprt} that $T_n=\inf\{k\ge 1: S_k\not\in(-A_n,B_n)\}$ and that
 \begin{align}\label{eqn:sprt}
d_{n,k}(X_{1}^{k},M_{1}^{k})=\begin{cases}
0&S_{k}\ge B_n\\
1&S_{k}\le -A_n\\
*&\mbox{otherwise}.
\end{cases}
\end{align}

 In the following lemma, we collect the properties of $\{S_{k}\}_{k=1}^{\infty}$ used in the proof of the achievability parts of Theorems~\ref{expectation} and \ref{probabilistic}.  These results may be of independent interest. For $i\in\{0,1\}$, let $U_{i}$ be the random variable with distribution $P_{\rho_{0},m_i^{*}}$ and let $W_{i}:=\log{P_{\rho_{0},m_i^{*}}(U_{i})}-\log{P_{\rho_{1},m_i^{*}}(U_{i})}$.
\begin{lemma}\label{lemma:sprt}
For the adaptive strategies $\{p_{k}\}_{k=1}^{\infty}$ defined in (\refeq{eqn:d1}) and (\refeq{eqn:d2}), the stochastic process  $\{S_{k}\}_{k=1}^{\infty}$ admits the following properties:
\begin{enumerate}[(i)]
\item\label{item:conditional} The conditional expectation of $Z_{j}$ given $\mathcal{F}_{j-1}$ is
\begin{align}\label{eqn:conditional}
\E_0[Z_j|\mathcal{F}_{j-1}]=\chi_{\{S_{j-1}\ge 0\}} D_{\M}(\rho_0\|\rho_1)+
 \chi_{\{S_{j-1}< 0\}} D(P_{\rho_0,m_1^*}\|P_{\rho_1,m_1^*}).
\end{align}
\item\label{item:partialsum} For sufficiently small $\lambda>0$, there exists $0<c<1$ such that
\begin{align}\label{eqn:partialsum}
\E_{0}[e^{-\lambda S_k}]\le c^{k}
\quad
\mbox{and}
\quad
\P_{0}(S_k<0)\le c^{k}.
\end{align} 
\item\label{item:bound}  Let $\hat{T}_{n}=\inf\{k\ge 1:S_{k}\ge B_{n}\}$. Then there exists some finite constant $C_{1}$ such that 
\begin{align}
-\E_0\big[S_{\hat{T}_n}-\hat{T}_nD_\M(\rho_0\|\rho_1)\big]\le C_1.
\end{align}
\item\label{item:expectation}  The limit of the normalized expectation of $S_{k}$ satisfies
\begin{align}\label{lemma:expectation}
\lim_{k\to\infty}\frac{\E_0[S_k]}{k}=\E[W_{0}]=D_{\M}(\rho_0\|\rho_1).
\end{align}
\item\label{item:variance} The limit of the normalized variance of $S_{k}$ is such that
\begin{align}\label{lemma:variance}
\lim_{k\to\infty}\frac{\E_0[(S_k-\E_{0}[S_{k}])^{2}]}{k}=\mathrm{Var}(W_{0}).
\end{align}
\end{enumerate}
\end{lemma}
In particular, Part~(\refeq{item:bound}) is crucial for the achievability proof of Theorem \ref{expectation}. It says that under the hypothesis that $\rho=\rho_{0}$, the stochastic process $\big\{ \hat{T}_nD_\M(\rho_0\|\rho_1)-S_{\hat{T}_n}\big\}_{n=1}^\infty$ is bounded above in expectation under $\E_0$ for all $n \in\mathbb{N}$. 
\begin{proof} We first prove part~(\refeq{item:conditional}). Note that
\begin{align}
\E_0[Z_j|\mathcal{F}_{j-1}]&=\E_0[Z_j|S_{j-1}]\\
&=\chi_{\{S_{j-1}\ge 0\}}\E_0[Z_{j}|S_{j-1}\ge 0]+\chi_{\{S_{j-1}< 0\}}\E_0[Z_{j}|S_{j-1}< 0]\\
&=\chi_{\{S_{j-1}\ge 0\}}\E_0\left[\log\frac{P_{\rho_0,m_0^*}(X_j)}{P_{\rho_1,m_0^*}(X_j)}\bigg|S_{j-1}\ge 0\right]\notag\\*
&\hspace{2.4cm}+\chi_{\{S_{j-1}< 0\}}\E_0\left[\log\frac{P_{\rho_0,m_1^*}(X_j)}{P_{\rho_1,m_1^*}(X_j)}\bigg|S_{j-1}< 0\right]\label{eqn:conditional1}\\
&=\chi_{\{S_{j-1}\ge 0\}} D_{\M}(\rho_0\|\rho_1)+
 \chi_{\{S_{j-1}< 0\}} D(P_{\rho_0,m_1^*}\|P_{\rho_1,m_1^*}).\label{eqn:conditional2}
\end{align}
where~(\refeq{eqn:conditional1}) follows from~(\refeq{dynamics}) and (\refeq{eqn:conditional2}) follows from the definition of~$m_0^*$. Therefore we have~(\refeq{eqn:conditional}) as desired.

Now we prove Part~(\refeq{item:partialsum}). Note that
\begin{align}
&\E_{0}[e^{-\lambda S_k}]=\E_0\Big[e^{-\lambda S_{k-1}}\E_0\big[e^{-\lambda Z_k}\big|\mathcal{F}_{k-1}\big]\Big]\\
&=\E_0\Big[e^{-\lambda S_{k-1}}\E_0\big[e^{-\lambda Z_k}\big|S_{k-1}\big]\Big]\\
&=\E_0\Big[e^{-\lambda S_{k-1}}\big(\chi_{\{S_{k-1> 0}\}}\E_0\big[e^{-\lambda Z_k}\big|M_{k}\!=\!m_0^*\big]\!+\!\chi_{\{S_{k-1\le 0}\}}\E_0\big[e^{-\lambda Z_k}\big|M_{k}\!=\!m_1^*\big]\big)\Big]\\
&\le \E_0\bigg[e^{-\lambda S_{k-1}}\max\Big\{\E_0\big[e^{-\lambda Z_k}\big|M_{k}=m_0^*\big],\E_0\big[e^{-\lambda Z_k}\big|M_{k}=m_1^*\big]\Big\}\bigg]\\
&\le \prod_{j=1}^{k}\max\Big\{\E_0\big[e^{-\lambda Z_j}\big|M_{j}=m_0^*\big],\E_0\big[e^{-\lambda Z_j}\big|M_{j}=m_1^*\big]\Big\}\\
&\le \Big(\max_{i=0,1}\E_0\big[e^{-\lambda W_i}\big]\Big)^k,\label{eqn:W1}
\end{align}
where~(\refeq{eqn:W1}) follows from the fact that the distribution of $W_i$ is the same as the conditional distribution of $Z_j$ given $M_{j}=m_i^*$. For sufficiently small $\lambda>0$, there exists some constant $c<1$ such that 
\begin{align}
\max_{i=0,1}\E_0\big[e^{-\lambda W_i}\big]<c,
\end{align}
which implies that $\E_{0}[e^{-\lambda S_k}]<c^{k}$. Using Markov's inequality, we then have that 
\begin{align}
\P_{0}(S_k<0)=\P_{0}(e^{-\lambda S_k}\ge 1)\le \E_{0}[e^{-\lambda S_k}]<c^{k}.
\end{align} This  completes the proof of Part~(\refeq{item:partialsum}).

Now we prove Part~(\refeq{item:bound}). Let $Y_k:=-(S_k-\sum_{j=1}^{k}\E_0[Z_j|\mathcal{F}_{j-1}])$. We first check that $\{Y_k\}_{k=1}^{\infty}$ is a martingale with respect to $\{\mathcal{F}_{k}\}_{k=1}^{\infty}$. Note that
\begin{align}
\E_0[Y_k|\mathcal{F}_{k-1}]&=-\E_0\Big[S_k-\sum_{j=1}^{k}\E_0[Z_j|\mathcal{F}_{j-1}]\Big|\mathcal{F}_{k-1}\Big]\\
&=-\E_0\Big[S_{k-1}-\sum_{j=1}^{k-1}\E_0[Z_j|\mathcal{F}_{j-1}]\bigg|\mathcal{F}_{k-1}\Big]\\
&=-\Big(S_{k-1}-\sum_{j=1}^{k-1}\E_0[Z_j|\mathcal{F}_{j-1}]\Big)\label{eqn:martingale}\\
&=Y_{k-1}
\end{align}
where (\refeq{eqn:martingale}) follows from the fact that $S_{k-1}-\sum_{j=1}^{k-1}\E_0[Z_j|\mathcal{F}_{j-1}]$ is $\mathcal{F}_{k-1}$-measurable.
Note that
\begin{align}
\E_0\big[|Y_k-Y_{k-1}|\big|\mathcal{F}_{k-1}\big]&=\E_0\big[|S_k-S_{k-1}-\E[Z_k|\mathcal{F}_{k-1}]|\big|\mathcal{F}_{k-1}\big]\\
&\le 2\E_{0}\big[|Z_k|\big|\mathcal{F}_{k-1}\big]\\
&\le 2C\label{eqn:aux2},
\end{align}
where (\refeq{eqn:aux2}) follows from  Lemma~\ref{lemma:mre}. 
Using Markov's inequality, we have that  for sufficiently small $\lambda>0$,
\begin{align}
\P_0(\hat{T}_n>k)&\le \P_0(S_k\le B_n)\\
&\le e^{\lambda B_n}\E_{0}[e^{-\lambda S_k}]\\
&\le e^{\lambda B_n}c^{k},\label{eqn:bpartialsum}
\end{align}
where~(\refeq{eqn:bpartialsum}) follows from~Part~(\refeq{item:partialsum}) of Lemma~\ref{lemma:sprt}. Therefore
\begin{align}
\E_0[\hat{T}_n]=\sum_{k=1}^{\infty}\P_{0}(\hat{T}_n\ge k)<\infty.\label{eqn:exp}
\end{align}
From~(\refeq{eqn:aux2}) and (\refeq{eqn:exp}), we see that the conditions of Theorem~\ref{optionalstopping} are satisfied for the stopping time  $\hat{T}_{n}$ and  the martingale $\{S_{k}-\sum_{j=1}^{k}\E_0[Z_j|\mathcal{F}_{j-1}]\}_{k=1}^{\infty}$. Therefore we have that
\begin{align}
-\E_0\bigg[S_{\hat{T}_n}-\sum_{j=1}^{\hat{T}_n}\E_0[Z_j|\mathcal{F}_{j-1}]\bigg]= -\E_0\big[Z_{1}-\E_0[Z_1|\mathcal{F}_{0}]\big]=0,
\end{align}
which further implies that
\begin{align}
&-\E_0\left[S_{\hat{T}_n}-\hat{T}_n D_{\M}(\rho_0\|\rho_1)]\right]= \E_0\bigg[\hat{T}_n D_{\M}(\rho_0\|\rho_1)-\sum_{j=1}^{\hat{T}_n}\E_0[Z_j|\mathcal{F}_{j-1}]\bigg]\\
&=\E_0\bigg[\sum_{j=1}^{\hat{T}_n} \big(D_{\M}(\rho_0\|\rho_1)-\chi_{\{S_{j-1}\ge 0\}} D_{\M}(\rho_0\|\rho_1)-
 \chi_{\{S_{j-1}< 0\}} D(P_{\rho_0,m_1^*}\|P_{\rho_1,m_1^*})\big)\bigg]\label{eqn:aux3}\\
&=\E_0\bigg[\sum_{j=1}^{\hat{T}_n} \chi_{\{S_{j-1}< 0\}}\big(D_{\M}(\rho_0\|\rho_1)-D(P_{\rho_0,m_1^*}\|P_{\rho_1,m_1^*})\big)\bigg]\label{eqn:aux31}\\
&\le \big(D_{\M}(\rho_0\|\rho_1)-D(P_{\rho_0,m_1^*}\|P_{\rho_1,m_1^*})\big)\E_0\bigg[\sum_{j=1}^{\infty}\chi_{\{S_{j-1}< 0\}}\bigg]\\
&=\big(D_{\M}(\rho_0\|\rho_1)-D(P_{\rho_0,m_1^*}\|P_{\rho_1,m_1^*})\big)\sum_{j=1}^{\infty}\P_0(S_{j-1}< 0),
\end{align}
where~(\refeq{eqn:aux3}) follows from~Part~(\refeq{item:conditional}) of Lemma~\ref{lemma:sprt}.
As $\P_0(S_{k}< 0)\le c^{k}$ from~Part~(\refeq{item:partialsum}) of Lemma~\ref{lemma:sprt}, we have that $\sum_{i=1}^{\infty}\P_0(S_{i-1}< 0)<\infty$.
We then complete the proof of Part~(\refeq{item:bound}) by setting
\begin{align}
C_1 :=\big(D_{\M}(\rho_0\|\rho_1)-D(P_{\rho_0,m_1^*}\|P_{\rho_1,m_1^*})\big)\sum_{i=1}^{\infty}\P_0(S_{i-1}< 0).
\end{align}

We now proceed to the proof of Part~(\refeq{item:expectation}).  From~Part(\refeq{item:conditional}) of Lemma~\ref{lemma:sprt}, it follows that for $j\ge 2$
\begin{align}\label{eqn:p1}
\E_0[Z_j]=\P_0(S_{j-1}\ge 0)D_{\M}(\rho_0\|\rho_1)+\P_0(S_{j-1}<0)D(P_{\rho_0,m_1^*}\|P_{\rho_1,m_1^*}), 
\end{align}
and
\begin{align}\label{eqn:probabilistic:aux0}
   \E_0[Z_j-\E_0[Z_j]|S_{j-1}] 
   &= \big(\chi_{\{S_{j-1}\ge 0\}}-\P_0(S_{j-1}\ge 0)\big)D_{\M}(\rho_0\|\rho_1)\notag\\
   &\quad
   +\big(\chi_{\{S_{j-1}< 0\}}-\P_0(S_{j-1}< 0)\big)D(P_{\rho_0,m_1^*}\|P_{\rho_1,m_1^*}).
\end{align}
Therefore we have that
\begin{align}\label{eqn:expectationparts}
\frac{\E_0[S_n]}{n}&=\frac{1}{n}\sum_{j=1}^{n}\left(\P_0(S_{j-1}\ge 0)D_{\M}(\rho_0\|\rho_1)+\P_0(S_{j-1}<0)D(P_{\rho_0,m_1^*}\|P_{\rho_1,m_1^*})\right).
\end{align}
As $\P_{0}(S_j\le 0)\le c^{j}$ for some $0<c<1$ from~Part~(\refeq{item:partialsum}) of Lemma~\ref{lemma:sprt}, we have that
\begin{align}
\lim_{j\to\infty}P_0(S_{j-1}\ge 0)D_{\M}(\rho_0\|\rho_1)&=D_{\M}(\rho_0\|\rho_1)\quad\mbox{and}\\
\lim_{j\to\infty}\P_0(S_{j-1}\!<\!0)D(P_{\rho_0,m_1^*}\|P_{\rho_1,m_1^*})&=0,
\end{align}
which together with~(\refeq{eqn:expectationparts}) implies that
\begin{align}\label{eqn:expectation}
&\lim_{n\to\infty}\frac{\E_0[S_n]}{n}\notag\\*
&=\lim_{n\to\infty} \frac{1}{n}\sum_{j=1}^{n}\P_0(S_{j-1}\!\ge\! 0)D_{\M}(\rho_0\|\rho_1)+\lim_{n\to\infty}\frac{1}{n}\sum_{j=1}^{n}\P_0(S_{j-1}\!<\!0)D(P_{\rho_0,m_1^*}\|P_{\rho_1,m_1^*})\\
&=D_{\M}(\rho_0\|\rho_1),
\end{align}
as desired.

We now proceed to the proof of Part (\refeq{item:variance}).  Note that
\begin{align}
&\E_0[(S_n-\E_0[S_n])^2]\notag\\
&=\E_0[(S_{n-1}-\E_0[S_{n-1}])^2]+\E_0[(Z_n-\E_0[Z_n])^2]+2\E_0[(S_n-\E_0[S_n])(Z_n-\E_0[Z_n])]\\
&=\sum_{j=1}^{n}\E_0[(Z_j-\E_0[Z_j])^2]+2\sum_{j=2}^{n}\E_0[(S_{j-1}-\E_0[S_{j-1}])(Z_j-\E_0[Z_j])],\label{eqn:ap:aux1}
\end{align}
where~(\refeq{eqn:ap:aux1}) follows by induction.

Using similar arguments as in the derivation of~(\refeq{eqn:expectation}), we have that
\begin{align}\label{eqn:variance}
\lim_{n\to\infty}\frac{1}{n}\sum_{j=1}^{n}\E_0[(Z_j-\E_0[Z_j])^2]= \E_0\left[\left(W_{0}-D_{\M}(\rho_0\|\rho_1)\right)^2\right].
\end{align}
From~Part~(\refeq{item:conditional}) of Lemma~\ref{lemma:sprt} we have that
\begin{align}
&\E_0\big[(S_{j-1}-\E_0[S_{j-1}])(Z_j-\E_0[Z_j])\big]\notag\\
&=\E_0\big[S_{j-1}\E_{0}[Z_j-\E_0[Z_j]|\mathcal{F}_{j-1}]\big]\\
&=\E_0\big[S_{j-1}\big\{(\chi_{\{S_{j-1}\ge 0\}}-\P_0(S_{j-1}\ge 0)\big)D_{\M}(\rho_0\|\rho_1)\notag\\
&\hspace{5cm}+\big(\chi_{\{S_{j-1}< 0\}}-\P_0(S_{j-1}< 0)\big)D(P_{\rho_0,m_1^*}\|P_{\rho_1,m_1^*})\big\}\big]\\
&=\underbrace{\E_0\big[S_{j-1}\big(1-\P_0(S_{j-1}\ge 0)\big)D_{\M}(\rho_0\|\rho_1)\chi_{\{S_{j-1}\ge 0\}}\big]}_{=: L_{1}}\notag\\
&\hspace{10mm}\underbrace{-\E_0\big[S_{j-1}\P_0(S_{j-1}\ge 0)D_{\M}(\rho_0\|\rho_1)\chi_{\{S_{j-1}< 0\}}\big]}_{=: L_{2}}\notag\\
&\hspace{15mm}\underbrace{-\E_0\big[S_{j-1}\P_0(S_{j-1}< 0)D(P_{\rho_0,m_1^*}\|P_{\rho_1,m_1^*})\chi_{\{S_{j-1}\ge 0\}}\big]}_{=: L_{3}}\notag\\
&\hspace{20mm}+\underbrace{\E_0\big[S_{j-1}\big(1-\P_0(S_{j-1}< 0)\big)D(P_{\rho_0,m_1^*}\|P_{\rho_1,m_1^*})\chi_{\{S_{j-1}< 0\}}\big]}_{=: L_{4}}\label{eqn:probabilistic:aux3}.
\end{align}
Now we bound the four terms $L_{1},L_{2},L_{3}$, and $L_{4}$ in~(\refeq{eqn:probabilistic:aux3}). Note that
\begin{align}
L_{1}&=\P_0(S_{j-1}
< 0)D_{\M}(\rho_0\|\rho_1)\E_0[|S_{j-1}|]\le CD_{\M}(\rho_0\|\rho_1)(j-1)c^{j-1}\label{eqn:p2},
\end{align}
and 
\begin{align}
L_{2}&\le D_{\M}(\rho_0\|\rho_1)\E_0[-S_{j-1}\chi_{\{S_{j-1}< 0\}}]\le C D_{\M}(\rho_0\|\rho_1)(j-1)c^{j-1},\label{eqn:p3}
\end{align}
where (\refeq{eqn:p2}) and (\refeq{eqn:p3}) both follow from  Lemma~\ref{lemma:mre} that $|S_{j}|\le jC$ and Part~(\refeq{item:partialsum}) of Lemma~\ref{lemma:sprt} that $\P_0(S_{j-1}< 0)\le c^{j-1}$. Similarly we have that
\begin{align}
L_{3}&\le C D(P_{\rho_0,m_1^*}\|P_{\rho_1,m_1^*})(j-1)c^{j-1}\label{eqn:p4}
\end{align}
and 
\begin{align}
L_{4}&\le C D(P_{\rho_0,m_1^*}\|P_{\rho_1,m_1^*})(j-1)c^{j-1}.\label{eqn:p5}
\end{align}
Combining~(\refeq{eqn:probabilistic:aux3}),~(\refeq{eqn:p2}), (\refeq{eqn:p3}), (\refeq{eqn:p4}), and (\refeq{eqn:p5}), we have that
\begin{align}
&\sum_{j=2}^{n}\E_0[(S_{j-1}-\E_0[S_{j-1}])(Z_j-\E_0[Z_j])]  \notag\\*
&\le 2C D_{\M}(\rho_0\|\rho_1)\sum_{j=2}^{n}(j-1)c^{j-1}+2C D(P_{\rho_0,m_1^*}\|P_{\rho_1,m_1^*})\sum_{j=2}^{n}(j-1)c^{j-1}\\
&\le C_2\label{eqn:p6}
\end{align}
for some finite constant $C_2$. Combining~\eqref{eqn:variance} and~\eqref{eqn:p6}, we have~(\refeq{lemma:variance}) as desired.  This completes the proof of Lemma~\ref{lemma:sprt}.
\end{proof}

We prove the achievability part of Theorem~\ref{expectation} by showing that the sequence of SQPRTs $\big\{\big(\cX,\{p_{k},d_{n,k}\}_{k=1}^{\infty},T_{n}\big)\big\}_{n=1}^{\infty}$ with the sequence of parameters $\{(A_n,B_n)\}_{n=1}^{\infty}$ satisfy~(\refeq{eqn:R0}),~(\refeq{eqn:R1}), and~(\refeq{eqn:exp_con}).

First we upper bound the two types of error probabilities for  the SQPRT $\mathcal{S}_{n}$ with parameters~$A_{n}$ and $B_{n}$.  Note that
\begin{align}
\alpha_n&=\P_0(d_{T_{n}}=1)\\
&=\E_0[\chi_{\{S_{T_n}\le -A_n\}}]\\
&=\E_1[e^{S_{T_n}}\chi_{\{S_{T_n}\le -A_n\}}]\label{eqn:com}\\
&\le e^{-A_n},
\end{align}
where (\refeq{eqn:com}) follows from~Part~(\refeq{com:sht}) of Corollary~\ref{cor:sht}. Therefore we have that
\begin{align}
\alpha_n\le e^{-A_n}\quad\mbox{and}\quad \beta_n\le e^{-B_n}.
\end{align}
We now show that the sequence of SQPRTs $\big\{\mathcal{S}_{n}\big\}_{n=1}^{\infty}$ with parameters $\{(A_n,B_n)\}_{n=1}^{\infty}$ satisfies
the expectation constraint~(\refeq{eqn:exp_con}). 
Recall that from~Lemma~\ref{lemma:mre}  we have that 
\begin{align}\label{eqn:boundlog}
|Z_k|=\big|\log{\Tr[\rho_0 M_{k}(X_k)]}-\log{\Tr[\rho_i M_k(X_k)]}\big|\le C.
\end{align} 
 Let $\hat{T}_n=\inf\{k:S_k\ge B_n\}$. Then $T_n\le \hat{T}_n$ and 
\begin{align}
\E_0[T_n]&\le \E_0[\hat{T}_n]\\
&= \frac{-\E_0[S_{\hat{T}_n}-\hat{T}_nD_\M(\rho_0\|\rho_1)]+\E_0[S_{\hat{T}_n}]}{D_\M(\rho_0\|\rho_1)}\\
&\le \frac{-\E_0[S_{\hat{T}_n}-\hat{T}_nD_\M(\rho_0\|\rho_1)]+B_n+C}{D_\M(\rho_0\|\rho_1)}\label{stopped}\\
&\le \frac{C_1+B_n+C }{D_\M(\rho_0\|\rho_1)}\label{optional},
\end{align}
where $(\refeq{stopped})$ follows from (\refeq{eqn:boundlog}) and the fact that 
$S_{\hat{T}_n-1}\le B_n$, and (\refeq{optional}) follows from Part (i) of Lemma~\ref{lemma:sprt}.
Similarly, we have that
\begin{align}
\E_1[T_n]\le \frac{A_n+C_3}{D_\M(\rho_1\|\rho_0)}.
\end{align}
for some finite constant $C_{3}$.
From the definitions of $A_n$ and $B_n$, we conclude that there exists an integer $N$ such that for all $n>N$,
$\max\{\E_0[T_n],\E_1[T_n]\}<n$. Therefore the expectation constraint~(\refeq{eqn:exp_con}) is satisfied for the sequence of SQPRTs $\big\{\big(\cX,\{p_{k},d_{n,k}\}_{k=1}^{\infty},T_{n}\big)\big\}_{n=1}^{\infty}$ with the sequence of parameters $\{(A_n,B_n)\}_{n=1}^{\infty}$, which together with the upper bounds on $\alpha_n$ and $\beta_n$ implies that 
$(D_\M(\rho_1\|\rho_0)-\tau,D_\M(\rho_0\|\rho_1)-\tau)$ is an achievable error exponent pair. Due to the arbitrariness of $\tau>0$, we conclude that $\big(D_\M(\rho_1\|\rho_0),D_\M(\rho_0\|\rho_1)\big)$ is achievable. 

We now prove the achievability part of Theorem~\ref{probabilistic}  by showing that the sequence of SQPRTs $\big\{\mathcal{S}_{n}\big\}_{n=1}^{\infty}$ with the sequence of parameters $\{(A_n,B_n)\}_{n=1}^{\infty}$ satisfies~(\refeq{eqn:R0}),~(\refeq{eqn:R1}), and~(\refeq{eqn:prob_con}). Similar to the proof of the achievability part of Theorem~\ref{expectation}, we can show that
\begin{align}
\alpha_n\le e^{-A_n}\quad\mbox{and}\quad \beta_n\le e^{-B_n}.
\end{align} 
We now show that the sequence of SQPRTs $\big\{\mathcal{S}_{n}\big\}_{n=1}^{\infty}$ with parameters $\{(A_n,B_n)\}_{n=1}^{\infty}$ satisfies the probabilistic constraint~(\refeq{eqn:prob_con}). Let $\hat{T}_n=\inf\{k\ge 1:S_k\ge B_n\}$. Next consider, 
\begin{align}
\P_0(T_n\ge n)&\le \P_0(\hat{T}_n\ge n)\\
&\le \P_0(S_n<B_n)\\
&\le \frac{\E_0\big[\big(S_n-nD_{\M}(\rho_0\|\rho_1)\big)^2\big]}{n^2\tau^2}\label{eqn:p0}\\
&\le \frac{2\E_0[(S_n-\E_0[S_n])^2]+2\big(\E_0[S_n]-nD_{\M}(\rho_0\|\rho_1)\big)^2}{n^2\tau^2}\label{eqn:conversep1}\\
&\to 0\label{eqn:conversep2},
\end{align}
where (\refeq{eqn:p0}) follows from Chebyshev's inequality and~(\refeq{eqn:conversep2}) follows from Parts (\refeq{item:expectation}) and (\refeq{item:variance}) of Lemma~\ref{lemma:sprt}.
 Similarly, we have that $\P_1(T_n\ge n)\to 0$ as $n\to\infty$. Therefore, our proposed sequence of SQPRTs $\big\{\big(\cX,\{p_{k},d_{n,k}\}_{k=1}^{\infty},T_{n}\big)\big\}_{n=1}^{\infty}$ with the sequence of parameters $\{(A_n,B_n)\}_{n=1}^{\infty}$ satisfies the probabilistic constraint~(\refeq{eqn:prob_con}). Hence,  we conclude that 
$(D_\M(\rho_1\|\rho_0)-\tau,D_\M(\rho_0\|\rho_1)-\tau)$ is an achievable error exponent pair. Due to the arbitrariness of $\tau>0$, we conclude that $\big(D_\M(\rho_1\|\rho_0),D_\M(\rho_0\|\rho_1)\big)$ is achievable. 

\subsection{Converse Proofs of Theorems~\ref{expectation} and~\ref{probabilistic}.}
We first prove the converse of Theorem~\ref{expectation}.  The following lemma provides lower bounds on the error probabilities for a general SQHT $\big(\cX,\{\mu_k,d_{k}\}_{k=1}^{\infty},T\big)$.
\begin{lemma}\label{lemma:exp:ub}
For any SQHT $\big(\cX,\{\mu_k,d_{k}\}_{k=1}^{\infty},T\big)$ with adaptive strategies such that 
\begin{align}
\max_{i=0,1}\E_{i}[T]<\infty,
\end{align}
the following inequalities hold,
\begin{align}
\log\frac{1}{\beta}&\le \frac{\E_0[T] D_{\M}(\rho_{0}\|\rho_1)+1}{1-\alpha}\label{eqn:exp:type2}\quad\mbox{and}\\
\log\frac{1}{\alpha}&\le \frac{\E_1[T] D_{\M}(\rho_{1}\|\rho_0)+1}{1-\beta}.\label{eqn:exp:type1}
\end{align}
\end{lemma}
Let $\{\mathcal{S}_{n}\}_{n=1}^{\infty}$ be a sequence of SQHTs with adaptive strategies such that $\alpha_n\to 0$ and $\beta_n\to 0$ and the sequence $\{T_n\}_{n=1}^{\infty}$ satisfies the expectation constraint~(\refeq{eqn:exp_con}). Then from~(\refeq{eqn:exp:type2}) and~(\refeq{eqn:exp:type1}) in Lemma~\ref{lemma:exp:ub}, we have that
\begin{align}\label{eqn:exp1}
\limsup_{n\to\infty}\frac{1}{n}\log\frac{1}{\beta_n}&\le \limsup_{n\to\infty}\frac{\E_{n,0}[T_{n}] D_{\M}(\rho_{0}\|\rho_1)+1}{n(1-\alpha_{n})} \le D_{\M}(\rho_{0}\|\rho_1).
\end{align}
and
\begin{align}\label{eqn:exp2}
\limsup_{n\to\infty}\frac{1}{n}\log\frac{1}{\alpha_n}\le \limsup_{n\to\infty}\frac{\E_{n,1}[T_{n}]D_{\M}(\rho_{1}\|\rho_0)+1}{n(1-\beta_{n})}\le D_{\M}(\rho_{1}\|\rho_0).
\end{align}
We then conclude that any achievable error exponent pair $(R_0,R_1)$ is such that $R_0\le D_{\M}(\rho_{1}\|\rho_0)$ and $R_1\le D_{\M}(\rho_{0}\|\rho_1)$. Thus to complete the proof of the converse part of Theorem~\ref{expectation}, it suffices to prove Lemma~\ref{lemma:exp:ub}. 

\begin{proof}[Proof of Lemma~\ref{lemma:exp:ub}]
Recall from Section~\ref{sec:model} that $\P_i$ is the probability measure on $(\Omega,\mathcal{F})$ when the underlying state is $\rho_i$. Let $\mathcal{F}_{T}$ be the sub-$\sigma$-algebra generated by $T$ and let $\P_{i,T}$ be the restriction of $\P_i$ to the $\sigma$-algebra $\mathcal{F}_{T}$. Then $\exp(S_{T})$ is the Radon-Nikodym derivative of $\P_{0,T}$ with respect to $\P_{1,T}$. Thus $\E_0[S_{T}]$ is the relative entropy between $\P_{0,T}$ and $\P_{1,T}$. We define a stochastic kernel  $V$ with input alphabet $\Omega$  (with elements $\omega$) and output alphabet $\{0,1\}$ as follows:
\begin{align}
V(0|\omega):=\begin{cases}1&\mbox{if $d_{T}(\omega)=0$}\\ 0&\mbox{if $d_{T}(\omega)=1$}\end{cases}.
\end{align} Note that $\big(\P_0(d_{T}=0), \P_0(d_{T}=1)\big)$ is the probability vector of the output of $V$ when $\P_{0,T}$ is the input probability measure on $\Omega$. Similarly, $\big(\P_1(d_{T}=0), \P_1(d_{T}=1)\big)$ is the probability vector of the output of the channel $V$ when $\P_{1,T}$ is the input probability measure on $\Omega$. Then applying the data processing inequality to the classical relative entropy when $\big(\P_{0, T}, \P_{1, T}\big)$ is processed via the stochastic kernel $V$, we obtain, 
\begin{align}
D(\alpha\|1-\beta)=D\big(\P_0(d_{T}=1)\|\P_1(d_{T}=1)\big)&\le D(\P_{0,T}\|\P_{1,T})= \E_0[S_{T}]\label{type1dpi}.
\end{align}
where  the {\em binary relative entropy} is defined as $D(a\|b)=a\log\frac{a}{b}+(1-a)\log\frac{1-a}{1-b}$ for any $0\le a,b\le 1$. Similarly we have that
\begin{align}
D(\beta\|1-\alpha)&\le \E_1[-S_{T}]\label{type2dpi}.
\end{align}
Let $H(a)=-a\log a-(1-a)\log(1-a)$  be the {\em binary entropy function}. Then since $D(\alpha\|1-\beta)=-H(\alpha)+\alpha\log\big(\frac{1}{1-\beta}\big)+(1-\alpha)\log\big(\frac{1}{\beta}\big)$, it follows from~(\refeq{type1dpi}) that
\begin{align}\label{eqn:expectation:aux0}
    \log\frac{1}{\beta}&\le \frac{D(\alpha\|1-\beta)+1}{1-\alpha}\le \frac{\E_0[S_{T}]+1}{1-\alpha}.
\end{align}
From Part (i) of Lemma~\ref{lemma:summartingale} it follows that $\{S_k-kD_{\M}(\rho_{0}\|\rho_1)\}_{k=1}^{\infty}$ is a supermartingale. 
Then applying Theorem~\ref{optionalstopping} to the supermartingale $\{S_k-kD_{\M}(\rho_{0}\|\rho_1)\}_{k=1}^{\infty}$ and the stopping time $T$, we obtain
\begin{align}\label{eqn:expectation:aux1}
   \E_0[S_{T}-TD_{\M}(\rho_{0}\|\rho_1)] \le \E_0[S_{1}-D_{\M}(\rho_{0}\|\rho_1)]\le 0.
\end{align}
Combining~$(\refeq{eqn:expectation:aux0})$ and~(\refeq{eqn:expectation:aux1}), we obtain
\begin{align}
\log\frac{1}{\beta}&\le \frac{\E_0[S_{T}]+1}{1-\alpha}\le \frac{\E_0[T] D_{\M}(\rho_{0}\|\rho_1)+1}{1-\alpha}.\label{eqn:exp:prtype2}
\end{align}
Similarly, we have that
\begin{align}
\log\frac{1}{\alpha}&\le \frac{\E_1[T] D_{\M}(\rho_{1}\|\rho_0)+1}{1-\beta}.\label{eqn:exp:prtype1}
\end{align}
This completes the proof of Lemma~\ref{lemma:exp:ub} and therefore also concludes the proof of the converse of Theorem~\ref{expectation}.
\end{proof}

Now we prove the converse part of Theorem~\ref{probabilistic}.
Let $\big\{\big(\cX_n,\{\mu_{n,k},d_{n,k}\}_{k=1}^{\infty},T_{n}\big)\big\}_{n=1}^{\infty}$ be a sequence of SQHTs such that $\max\{\alpha_n,\beta_n\}\to 0$ as $n\to\infty$ and the sequence $\{T_n\}_{n=1}^{\infty}$ satisfies the probabilistic constraint~(\refeq{eqn:prob_con}).

Let $\lambda_n:=n(D_{\M}(\rho_0\|\rho_1)+\tau)$ for some $\tau>0$. For any $\gamma>0$ such that $\gamma< (1-\eps)/2$, $\P_{n,0}\left(T_n>n\right)\le \eps+\gamma$ for sufficiently large $n$. Applying Part~(\refeq{probs_{k}:sht}) of Corollary~\ref{cor:sht} to the SQHT  $\big(\cX_n,\{p_{n,k},d_{n,k}\}_{k=1}^{\infty},T_{n}\big)$, we have that
\begin{align}
\P_{n,0}(d_{T_{n}}=0)-e^{\lambda_n}\P_{n,1}(d_{n,T_{n}}=0)&\le \P_{n,0}\left(S_{n,T_n}\ge \lambda_n\right)\\
&\le \P_{n,0}\left(S_{n,T_n}\ge \lambda_n, T_n\le n\right)+\P_{n,0}\left(T_n>n\right)\\
&\le \P_{n,0}\left(S_{n,T_n}\ge \lambda_n, T_n\le n\right)+\eps+\gamma\\
&\le \P_{n,0}\bigg(\max_{1\le k\le n}S_{n,k}\ge \lambda_n\bigg)+\eps+\gamma,
\end{align}
which implies that
\begin{align}
\log\frac{1}{\beta_n}&\le \lambda_n-\log\left[\P_{n,0}(d_{T_{n}}=0)-\P_{n,0}\left(\max_{1\le k\le n}S_{n,k}\ge \lambda_n\right)-\eps-\gamma\right]\\
&=\lambda_n-\log\left[1-\alpha_n-\P_{n,0}\left(\max_{1\le k\le n}S_{n,k}\ge \lambda_n\right)-\eps-\gamma\right].\label{type2p1}
\end{align}
We now upper bound $\P_{n,0}\left(\max_{1\le k\le n}S_{n,k}\ge \lambda_n\right)$. From Part (ii) of Lemma~\ref{lemma:summartingale} it follows that  $\{S_{n,k}\}_{k=1}^{\infty}$ is a submartingale, which together with Jensen's inequality, implies that  for any $t>0$, $\{\exp(tS_{n,k})\}_{k=1}^{\infty}$ is also a submartingale. For any POVM $m\in\M_{\mathcal{X}}$, let $Y_{m}$ be the random variable with probability mass function $P_{\rho_0,m}$ and let $W_{m}:=\log P_{\rho_0,m}(Y_{m})-\log P_{\rho_1,m}(Y_{m})$. Recall that $P_{\rho_0,m}(x)=\Tr[\rho_{0}m(x)]$. Using Theorem~\ref{doob}, we have that
\begin{align}
\P_{n,0}\left(\max_{1\le k\le n}S_{n,k}\ge \lambda_n\right)&\le e^{-t\lambda_n}\E_{n,0}\big[e^{tS_{n,n}}\big]\\
&=e^{-t\lambda_n}\E_{n,0}\bigg[e^{tS_{n,n-1}}\E_{n,0}\big[e^{tZ_n}\big|\mathcal{F}_{n-1}\big]\bigg]\\
&\le e^{-t\lambda_n}\E_{n,0}\bigg[e^{tS_{n,n-1}}\sup_{m\in\M_{\mathcal{X}}}\E_0\big[e^{tW_{m}}\big]\bigg]\\
&\le e^{-t\lambda_n}\bigg(\sup_{m\in\M_{\mathcal{X}}}\E_0\big[e^{tW_{m}}\big]\bigg)^n.\label{type2p2}
\end{align}
As $|W_{m}|\le C$ from Lemma~\ref{lemma:mre}, we can then apply Taylor's theorem to the function $t\mapsto \E_0 [ e^{tW_m} ]$ in a neighborhood of $0$ to obtain
\begin{align}
\sup_{m\in\M_{\mathcal{X}}}\E_0\big[e^{tW_{m}}\big]\le 1+tD_\M(\rho_0\|\rho_1)+C_4t^2
\end{align}
for some finite constant $C_4$, which together with~(\refeq{type2p2}) implies that for sufficiently small $t>0$ 
\begin{align}\label{eqn:probabilistic:aux4}
e^{-t(D_{\M}(\rho_0\|\rho_1)+\tau)}\sup_{m\in\M_{\mathcal{X}}}\E_0\big[e^{tW_{m}}\big]<1.
\end{align}
Then combing~(\refeq{type2p1}),~(\refeq{type2p2}) and~(\refeq{eqn:probabilistic:aux4}), we obtain
\begin{align}\label{type2p3}
\limsup_{n\to\infty}\frac{1}{n}\log\frac{1}{\beta_n}\le D_{\M}(\rho_{0}\|\rho_1)+\tau.
\end{align}
Using similar arguments as in the derivation of~(\refeq{type2p3}), we have that
\begin{align}\label{type1p1}
\limsup_{n\to\infty}\frac{1}{n}\log\frac{1}{\alpha_n}\le D_{\M}(\rho_{1}\|\rho_0)+\tau.
\end{align}
Due to the arbitrariness of $\tau>0$, we conclude that any achievable error exponent pair $(R_0,R_1)$ is such that $R_0\le D_{\M}(\rho_{1}\|\rho_0)$ and $R_1\le D_{\M}(\rho_{0}\|\rho_1)$, as desired. This completes the converse part of Theorem~\ref{probabilistic}.

\subsection{Achievability Proof of Theorem~\ref{non-adaptive}}

To prove that any pair  $(R_{0},R_{1})\in\overline{\mathrm{Conv}(\mathcal{C})}$ is achievable, it suffices to show that any pair  $(R_{0},R_{1})\in\mathrm{Conv}(\mathcal{C})$ is achievable. For any finite set $\mathcal{X}$ and any POVM $m\in\M_{\mathcal{X}}$, let 
\begin{align}
\mathcal{D}(m,\mathcal{X}):=\big\{\big(0,0\big),\big(0, D(P_{\rho_0,m}\|P_{\rho_{1}, m})\big),\big(D(P_{\rho_1,m}\|P_{\rho_0, m}), 0\big),\big(D(P_{\rho_1,m}\|P_{\rho_0, m}), D(P_{\rho_0,m}\|P_{\rho_{1}, m})\big)\big\},\notag
\end{align}
which is the set of corner points of the achievable error exponent region $\{(R_{0},R_{1}) \in \mathbb{R}_+^2: R_{i}\le D(P_{\rho_{1-i},m}\|P_{\rho_{i},m})\,\,\mbox{for $i=0,1$}\}.$ Let
$\mathcal{D}=\cup_{\mathcal{X}}\cup_{m\in\M_{\mathcal{X}}}\mathcal{D}(m,\mathcal{X}).$ Then we have $
\mathrm{Conv}(\mathcal{D})=\mathrm{Conv}(\mathcal{C})$. Therefore, the achievability of points in $\mathrm{Conv}(\mathcal{C})$ is equivalently to the achievability of points in $\mathrm{Conv}(\mathcal{D})$.
 Let $(R_{0},R_{1})$ be a point in $\mathrm{Conv}(\mathcal{C})$. Since the region of error exponent pairs is a subset of $\mathbb{R}_+^2$, then it follows from Carath\'eodory's theorem~\cite[Theorem 17.1, pp.~155]{Rockafellar} that there exists three points $\{(R_{0}^{(j)},R_{1}^{(j)})\}_{j=1}^{3}\subset \mathcal{D}$ such that 
\begin{align}
(R_{0},R_{1})=t_{1}(R_{0}^{(1)},R_{1}^{(1)})+t_{2}(R_{0}^{(2)},R_{1}^{(2)})+t_{3}(R_{0}^{(3)},R_{1}^{(3)}),
\end{align}
where $0\le t_{1},t_{2},t_{3}\le 1$ and $t_{1}+t_{2}+t_{3}=1$.

For $j\in\{1,2,3\}$, as $(R_{0}^{(j)},R_{1}^{(j)})\in \mathcal{D}$, there exists $(\mathcal{X}_{j},m_{j})$ be such that $(R_{0}^{(j)},R_{1}^{(j)})\in \mathcal{D}(\mathcal{X}_{j},m_{j})$. Then for $i\in\{0,1\}$
\begin{align}
R_{i}\le t_{1}D(P_{\rho_{1-i},m_{1}}\|P_{\rho_{i}, m_{1}})+t_{2}D(P_{\rho_{1-i},m_{2}}\|P_{\rho_{i}, m_{2}})+t_{3}D(P_{\rho_{1-i},m_{3}}\|P_{\rho_{i}, m_{3}}).
\end{align}
Thus if we can show that any convex combination of $\{(D(P_{\rho_{1},m_{j}}\|P_{\rho_0, m_{j}}),D(P_{\rho_0,m_{j}}\|P_{\rho_1, m_{j}}))\}_{j=1}^{3}$ is achievable, we can show that $(R_{0},R_{1})$ is also achievable.

We first consider case that $t_{1}=\frac{r_{1}}{q},\ t_{2}=\frac{r_{2}}{q}$ and $t_{3}=\frac{r_{3}}{q}$ for positive integers $r_{1},r_{2},r_{3}$ and $q$ such that $r_{1}+r_{2}+r_{3}=q$. Extrapolating this special case to the general case of irrational convex combinations can be done via standard approximation  arguments. Thus, we aim to show that 
\begin{align}\label{eqn:midpoint}
&\frac{1}{q}\bigg(\sum_{j=1}^{3}r_{j}D(P_{\rho_{1},m_{j}}\|P_{\rho_0, m_{j}}), \sum_{j=1}^{3}r_{j}D\big(P_{\rho_0,m_{j}}\big\|P_{\rho_1, m_{j}}\big)\bigg)
\end{align}
is achievable. Let 
$\mathcal{X}$ be the disjoint union of $\mathcal{X}_{1}$, $\mathcal{X}_{2}$, and $\mathcal{X}_{3}$. Then $m_{1}$, $m_{2}$ and $m_{3}$ are POVMs in $\M_\mathcal{X}$. We first define the non-adaptive strategies used in the SQPRTs. For any two integers $q_{1}$ and $q_{2}$, let $r(q_{1},q_{2})$ be the remainder of $q_{1}$ divided by $q_{2}$. Let 
\begin{align}
\mathcal{J}_{1}&:=\{k\in\mathbb{N}:1\le r(k, r_{1}+r_{2}+r_{3})\le r_{1}\},\\
\mathcal{J}_{2}&:=\{k\in\mathbb{N}:r_{1}+1\le r(k, r_{1}+r_{2}+r_{3})\le r_{1}+r_{2}\},\\
\mathcal{J}_{3}&:=\{k\in\mathbb{N}:r(k,r_{1}+r_{2}+r_{3})=0 \ \mbox{or}\ r_{1}+r_{2}+1\le r(k, r_{1}+r_{2}+r_{3})\le r_{1}+r_{2}+r_{3} \! -\! 1\}.
\end{align}
The POVM used on the $k$-th copy of the underlying state $\rho$ is $M_{k}=m_{j}$ if $k\in\mathcal{J}_{j}$ for $j\in\{1,2,3\}$. 

Note that when the underlying state is $\rho_i$, the sequence of  $\{X_k\}_{k=1}^{\infty}$ obtained from the POVMs $\{M_{k}\}_{k=1}^{\infty}$ applied to the underlying state is an independent sequence of random variables with
\begin{align}
 P_{\rho_{i},M_{k}}(X_k=x):=\begin{cases}\Tr\big[\rho_i m_{1}(x)\big]&k\in\mathcal{J}_{1}\\ \Tr\big[\rho_i m_{2}(x)\big]&k\in\mathcal{J}_{2}\\
 \Tr\big[\rho_i m_{3}(x)\big]&k\in\mathcal{J}_{3}\end{cases}
\end{align}
for any $x\in\mathcal{X}$.  Recall from~(\refeq{eqn:llr}) and~(\refeq{eqn:sk}) that $Z_{k}=\log{P_{\rho_{0},M_{k}}(X_{k})}-\log P_{\rho_1,M_{k}}(X_{k})$ and $S_{k}=\sum_{j=1}^{k}Z_{j}$. For any $0<\tau<\frac{1}{q}\min_{i=0,1}\big\{\sum_{j=1}^{3}r_{j}D\big(P_{\rho_{i},m_{j}}\|P_{\rho_{1-i}, m_{j}}\big) \big\}$, let
\begin{align}
A_n&:=\bigg(\frac{n}{q}\sum_{j=1}^{3}r_{j}D(P_{\rho_{1},m_{j}}\|P_{\rho_0, m_{j}})\bigg)-n\tau \quad\mbox{and}\quad B_n:=\bigg(\frac{n}{q}\sum_{j=1}^{3}r_{j}D(P_{\rho_0,m_{j}}\|P_{\rho_1, m_{j}})\bigg)-n\tau.
\end{align}
For any $n\ge 1$, let $\mathcal{S}_{n}=\big(\cX,\{p_{k},d_{n,k}\}_{k=1}^{\infty}, T_n\big)$ be the SQPRT with parameters $A_{n}$ and $B_{n}$. Recall that $T_n=\inf\{k\ge 1: S_k\not\in(-A_n,B_n)\}$ and that
 \begin{align}\label{eqn:na:sprt}
d_{n,k}(X_{1}^{k},M_{1}^{k})=\begin{cases}
0&S_{k}\ge B_n\\
1&S_{k}\le -A_n\\
*&\mbox{otherwise}.
\end{cases}
\end{align}  
Thus to complete the proof of the achievability of the error exponent pair defined in~(\refeq{eqn:midpoint}) under the expectation (resp. probabilistic) constraints, we only need to show that $\big\{\mathcal{S}_{n}\}_{n=1}^{\infty}$  satisfies~(\refeq{eqn:R0}),~(\refeq{eqn:R1}), and~(\refeq{eqn:exp_con}) (resp.~(\refeq{eqn:prob_con})). Using similar arguments as in the proof of achievability part of Theorem~\ref{expectation}, we have that
\begin{align}
\alpha_n\le e^{-B_n}\quad \mbox{and}\quad \beta_n\le e^{-A_n}.
\end{align}
 Due to the similarity of the proof of $\E_0[T_n]\le n$ (resp.\ $\P_0(T_n>n)< \eps$) and $\E_1[T_n]\le n$ (resp.\ $\P_1(T_n>n)< \eps$), we only prove  the former statements, i.e.,  $\E_0[T_n]\le n$ and $\P_0(T_n>n)< \eps$. 

Similarly as in the proof of Theorem~\ref{expectation}, we define $\hat{T}_n$ to be the first time that $S_k$ is larger than~$B_n$. We now prove $\E_0[T_n]\le n$ for sufficiently large $n$.  As $\{(M_{k},X_{k})\}_{k=1}^{\infty}$ is a sequence of independent random variables, it follows that $\{S_{k}-\E_{0}[S_{k}]\}_{k=1}^{\infty}$ is a martingale. From Part (iii) of Lemma~\ref{lemma:mre}, we have that $\left|Z_{k}\right|\le C.$
Then it follows from Theorem~\ref{optionalstopping} we have that
\begin{align}
\E_0\bigg[S_{\hat{T}_n}-\sum_{k=1}^{\hat{T}_n}\E_0[Z_k]\bigg]=0,
\end{align}
which further implies that
\begin{align}
\E_0[S_{\hat{T}_n}]&=\E_0\bigg[\sum_{k=1}^{\hat{T}_n}\E_0[Z_{k}]\bigg]\\
&\ge\E_0\bigg[\Big\lfloor\frac{\hat{T}_n}{q}\Big\rfloor\bigg]\sum_{j=1}^{3}r_{j}D\big(P_{\rho_{1},m_{j}}\big\|P_{\rho_0, m_{j}}\big)\\
&\ge\E_0\Big[\frac{\hat{T}_n}{q}-1\Big]\sum_{j=1}^{3}r_{j}D\big(P_{\rho_{1},m_{j}}\big\|P_{\rho_0, m_{j}}\big).\label{eqn:na:wald}
\end{align}
From the definition of 
$\hat{T}_n$, we have that 
\begin{align}
S_{\hat{T}_n}=S_{\hat{T}_n-1}+Z_{\hat{T}_n}\le B_n+C,
\end{align} which together with~(\refeq{eqn:na:wald}) further implies that
\begin{align}
    \E_0[T_n]&\le \E_0[\hat{T}_n]\\
    &\le \frac{q(B_n+C)}{\sum_{j=1}^3 r_{j}D\big(P_{\rho_{1},m_{j}}\big\|P_{\rho_0, m_{j}}\big) }+q \\
    &=\frac{n \big[\sum_{j=1}^3 r_{j}D\big(P_{\rho_{1},m_{j}}\big\|P_{\rho_0, m_{j}}\big)\big] -qn\tau+qC}{\sum_{j=1}^3 r_{j}D\big(P_{\rho_{1},m_{j}}\big\|P_{\rho_0, m_{j}}\big)}+q \\
    &<n,
\end{align}
for sufficiently large $n$. Therefore we complete the proof that the sequence of SQPRTs $\big\{\mathcal{S}_n\}_{n=1}^{\infty}$ satisfies the expectation constraint~(\refeq{eqn:exp_con}) as desired. 

Now we prove that $\P_0(T_n>n)< \eps$. For $j\in\{1,2,3\}$,  let $\mathcal{J}_{n}(j)=\{k\in \mathcal{J}_{j}:k\le n\}$ and let $S_{n}^{(j)}=\sum_{k\in\mathcal{J}_{n}(j)}Z_{j}$.  Note that
\begin{align}
    \P_0(T_n>n)&\le \P_0(\hat{T}_n>n) \\
    &\le \P_0(S_n< B_n) \\
    &= \P_0(S_{n}^{(1)}+S_{n}^{(2)}+S_{n}^{(3)}< B_n)\\
    &=\P_0\bigg(\sum_{j=1}^{3}S_{n}^{(j)}-\frac{ nr_{j}D\big(P_{\rho_0,m_{j}}\big\|P_{\rho_1, m_{j}}\big)}{q}< -n\tau\bigg) \\
    &\le\sum_{j=1}^{3}\P_0\bigg(S_{n}^{(j)}-\frac{ nr_{j}D\big(P_{\rho_0,m_{j}}\big\|P_{\rho_1, m_{j}}\big)}{q}< -\frac{ nr_{j}}{q}\tau\bigg),\label{eqn:na:union}
\end{align}
where~(\refeq{eqn:na:union}) follows from the union bound. For $j\in\{1,2,3\}$, we have that 
\begin{align}
&\P_0\bigg(S_{n}^{(j)}-\frac{ nr_{j}D\big(P_{\rho_0,m_{j}}\big\|P_{\rho_1, m_{j}}\big)}{q}< -\frac{ nr_{j}}{q}\tau\bigg)\notag\\
&=\P_0\bigg(S_{n}^{(j)}-|\mathcal{J}_{n}(j)|D\big(P_{\rho_0,m_{j}}\big\|P_{\rho_1, m_{j}}\big)\le  -\frac{ nr_{j}}{q}\tau+\Big(\frac{ nr_{j}}{q}-|\mathcal{J}_{n}(j)|\Big)D\big(P_{\rho_0,m_{j}}\big\|P_{\rho_1, m_{j}}\big)\bigg)\\
&\le \P_0\bigg(S_{n}^{(j)}-|\mathcal{J}_{n}(j)|D\big(P_{\rho_0,m_{j}}\big\|P_{\rho_1, m_{j}}\big)\le -|\mathcal{J}_{n}(j)|\tau+r_{j}\Big(\tau+D\big(P_{\rho_0,m_{j}}\big\|P_{\rho_1, m_{j}}\big)\Big)\bigg)\label{eqn:prob:rem},
\end{align}
where~(\refeq{eqn:prob:rem}) follows from the fact that $\big||\mathcal{J}_{n}(j)|-\frac{ nr_{j}}{q}\big|\le r_{j}$.  As the sequence of random variables $\{X_k:k\in\mathcal{J}_{j}\}$ for $j \in \{1,2,3\}$ is i.i.d.,   using the weak law of large numbers, we obtain
\begin{align}
&\lim_{n\to\infty}\P_0\bigg(S_{n}^{(j)}-\frac{ nr_{j}D\big(P_{\rho_0,m_{j}}\big\|P_{\rho_1, m_{j}}\big)}{q}< -\frac{ nr_{j}}{q}\tau\bigg)\notag\\
&\le \lim_{n\to\infty}\P_0\bigg(S_{n}^{(j)}-|\mathcal{J}_{n}(j)|D\big(P_{\rho_0,m_{j}}\big\|P_{\rho_1, m_{j}}\big)\notag\\
&\qquad\qquad\qquad\le -|\mathcal{J}_{n}(j)|D\big(P_{\rho_0,m_{j}}\big\|P_{\rho_1, m_{j}}\big)+r_{j}\Big(\tau+D\big(P_{\rho_0,m_{j}}\big\|P_{\rho_1, m_{j}}\big)\Big)\bigg)\\
&=0.\label{eqn:prob:jterm}
\end{align}
Combining~(\refeq{eqn:na:union}) and~(\refeq{eqn:prob:jterm}), we conclude that $\P_0(T_n>n)<\eps$ for sufficiently large $n$, which completes the proof that the sequence of SQPRTs $\big\{\mathcal{S}_n\}_{n=1}^{\infty}$ satisfies the probabilistic constraint~(\refeq{eqn:prob_con}) as desired.

\subsection{Converse Proof of Theorem~\ref{non-adaptive}}
  Now we prove the converse part under both the expectation or the probabilistic constraint. Before we present the proof, we provide an upper bound on the tail probability of the maximal sum for any SQHT with non-adaptive strategies.
 \begin{lemma}\label{lemma:na:maximalsum}
For any SQHT $\big(\cX,\{\mu_{k},d_{k}\}_{k=1}^{\infty},T\big)$ with non-adaptive strategies and any $\lambda>0$, we have that
\begin{align}
\P_0\left(\max_{1\le j\le k}S_j\ge \E_0[S_k]+\lambda\right) &\le \frac{\E_{0}[(S_k-\E_{0}[S_{k}])^{2}]}{\lambda^{2}}\label{im1}\quad\mbox{and}\\
\P_1\left(-\min_{1\le j\le k}S_j\ge \E_1[-S_k]+\lambda\right)&\le \frac{\E_{1}[(-S_k-\E_{1}[-S_{k}])^{2}]}{\lambda^{2}}.\label{eqn:tail:p1}
\end{align}
 \end{lemma}
  \begin{proof}[Proof of Lemma~\ref{lemma:na:maximalsum}]
 As $\{\mu_{k}\}_{k=1}^\infty$ is a sequence of non-adaptive strategies, the process $\{(X_{j},M_{j})\}_{j=1}^{\infty}$ is an independent sequence of random variables. Recall from~(\refeq{eqn:llr}) and~(\refeq{eqn:sk}) that $Z_{k}=\log P_{\rho_{0},M_{k}}(X_{k})-\log P_{\rho_{1},M_{k}}(X_{k})$ and $S_{k}=\sum_{j=1}^{k}Z_{j}$. Therefore, $\{Z_{k}\}_{k=1}^{\infty}$ is a sequence of independent random variables and thus $\{S_{k}-\E_{0}[S_{k}]\}_{k=1}^{\infty}$ is a martingale. Hence $\{(S_{k}-\E_{0}[S_{k}])^{2}\}_{k=1}^{\infty}$ is a submartingale. Then we have that for any $\lambda>0$,
\begin{align}
\P_0\left(\max_{1\le j\le k}S_j\ge \E_0[S_k]+\lambda\right)&\le \P_0\left(\max_{1\le j\le k}(S_j-\E_{0}[S_{j}])\ge \lambda\right)\label{eqn:Chebyshev1}\\
&\le  \P_0\left(\max_{1\le j\le k}(S_j-\E_{0}[S_{j}])^{2}\ge \lambda^{2}\right)
\label{eqn:na:aux0}\\
&\le \frac{\E_{0}[(S_k-\E_{0}[S_{k}])^{2}]}{\lambda^{2}},\label{im_1}
\end{align}
where~(\refeq{eqn:Chebyshev1}) follows from the fact that $0\le \E_{0}[S_{j}]\le \E_{0}[S_{k}]$ and~(\refeq{im_1}) follows from Theorem~\ref{doob}.
Similarly, we have~(\refeq{eqn:tail:p1}). This completes the proof of Lemma~\ref{lemma:na:maximalsum}.
\end{proof}

We start by proving the converse of Theorem~\ref{non-adaptive} under the  expectation constraint. Let $(R_{0},R_{1})$ be an achievable error exponent pair. Suppose $\big\{\big(\cX_n,\{\mu_{n,k},d_{n,k}\}_{k=1}^{\infty},T_{n}\big)\big\}_{n=1}^{\infty}$ is a sequence of SQHTs with non-adaptive strategies satisfying~(\refeq{eqn:R0}),~(\refeq{eqn:R1}), and~(\refeq{eqn:exp_con}).  
For any fixed   $\gamma$ such that $0<\gamma< \min\{R_{0},R_{1}\}$, we have from Definition~\ref{def:ach_ee} that
\begin{align}
\max\{\E_{n,0}[T_{n}],\E_{n,1}[T_{n}]\}&\le n+\gamma,\label{eqn:nae:exp}\\
R_{0}-\gamma&\le \frac{1}{n}\log\frac{1}{\alpha_{n}},\label{eqn:nae:r0}\\
R_{1}-\gamma&\le  \frac{1}{n}\log\frac{1}{\beta_{n}},\label{eqn:nae:r1}
\end{align}
 for sufficiently large $n$. Recall from~(\refeq{def:znj}) and~\eqref{def:snk} that $Z_{n,j}=\log P_{\rho_{0},M_{n,j}}(X_{n,j})-\log P_{\rho_{1},M_{n,j}}(X_{n,j})$ and $S_{n,k}=\sum_{j=1}^{k}Z_{n,j}$.  Let $\lambda_{n}=\E_{n,0}[S_{n,n}]+n\gamma$. 
 Then from Part~(\refeq{probs_{k}:sht}) of Corollary~\ref{cor:sht}, it follows that
\begin{align}\label{eqn:st:lb}
1-\alpha_{n}-e^{\lambda_{n}}\beta_{n}&\le \P_{n,0}\Big(S_{n,T_{n}}\ge \lambda_{n}\Big)\\
&\le \P_{n,0}\Big(S_{n,T_{n}}\ge \lambda_{n}, T_{n}\le n+\sqrt{n}\Big)+\P_{n,0}\Big(T_{n}\ge n+\sqrt{n}\Big)\\
&\le \P_{n,0}\Big(\max_{1\le k\le n+\sqrt{n}}S_{n,k}\ge \lambda_{n}\Big)+\P_{n,0}\Big(T_{n}\ge n+\sqrt{n}\Big).\label{eqn:nae:lb}
\end{align}
Note that for sufficiently large $n$, by Markov's inequality, 
\begin{align}
\P_{n,0}\Big(T_{n}\ge n+\sqrt{n}\Big)&\le \frac{\E_{n,0}[T_{n}]}{n+\sqrt{n}}\\
&\le  \frac{n+\gamma}{n+\sqrt{n}}\label{eqn:sqrt}\\
&\le 1-\frac{1}{3\sqrt{n}}\label{eqn:sqrt1},
\end{align}
where~(\refeq{eqn:sqrt}) follows from~(\refeq{eqn:nae:exp}). Combining~(\refeq{eqn:nae:lb}) and~(\refeq{eqn:sqrt1}), we have that
\begin{align}
\frac{1}{n}\log\frac{1}{\beta_{n}}\le  \frac{\E_{n,0}[S_{n,n}]}{n}+\gamma-\frac{1}{n}\log\bigg(\frac{1}{3\sqrt{n}}-\alpha_{n}-\P_{n,0}\Big(\max_{1\le k\le n+\sqrt{n}}S_{n,k}\ge \lambda_{n}\Big)\bigg)\label{eqn:nae:type2}.
\end{align}
Since\footnote{More precisely, $n+\sqrt{n}$  (in the subscripts of $S_{n, \cdot }$ and the sums) should be replaced by $n + \lfloor \sqrt{n} \rfloor$ but we omit the floor operators to avoid notational clutter.}
\begin{align}\label{eqn:nae:lb:var}
\lambda_{n}-\E_{n,0}[S_{n,n+\sqrt{n}}]= n\gamma-\sum_{j=n+1}^{n+\sqrt{n}}\E_{n,0}[Z_{n,j}]\ge n\gamma-\sqrt{n} C>0
\end{align}
for sufficiently large $n$, then we have  
\begin{align}
\P_{n,0}\Big(\max_{1\le k\le n+\sqrt{n}}S_{n,k}\ge \lambda_{n}\Big)&=\P_{n,0}\Big(\max_{1\le k\le n+\sqrt{n}}S_{n,k}\ge \E_{n,0}[S_{n,n+\sqrt{n}}]+\lambda_{n}-\E_{n,0}[S_{n,n+\sqrt{n}}]\Big)\\
&\le \frac{\E_{n,0}[(S_{n,n+\sqrt{n}}-\E_{n,0}[S_{n,n+\sqrt{n}}])^{2}]}{(\lambda_{n}-\E_{n,0}[S_{n,n+\sqrt{n}}])^{2}}\label{eqn:na:ptail0}\\
&\le \frac{1}{(\lambda_{n}-\E_{0}[S_{n,n+\sqrt{n}}])^{2}}\sum_{j=1}^{n+\sqrt{n}}\E_{n,0}[Z_{n,j}^{2}]\label{eqn:independence}\\
&\le \frac{(n+\sqrt{n})C^{2}}{(n\gamma-\sqrt{n}C)^{2}},\label{eqn:na:etail2}
\end{align}
where~(\refeq{eqn:na:ptail0}) follows from~(\refeq{im1}) in Lemma~\ref{lemma:na:maximalsum},~(\refeq{eqn:independence}) follows from the fact that $\{Z_{n,j}\}_{j=1}^{\infty}$ is an independent sequence of random variables for any fixed $n$, and~(\refeq{eqn:na:etail2}) follows from  Lemma~\ref{lemma:mre}, namely that $|Z_{n,j}|\le C$. 
Then from~(\refeq{eqn:na:etail2}), we have that for sufficiently large $n$,
\begin{align}
\frac{1}{3\sqrt{n}}-\alpha_{n}-\P_{n,0}\Big(\max_{1\le k\le n+\sqrt{n}}S_{n,k}\ge \lambda_{n}\Big)&\ge \frac{1}{3\sqrt{n}}-e^{-n(R_{0} - \gamma)}-\frac{(n+\sqrt{n})C^{2}}{(n\gamma-\sqrt{n}C)^{2}}\\
&\ge\frac{1}{4\sqrt{n}},
\end{align}
which implies that
\begin{align}
- \frac{1}{n}\log\bigg(\frac{1}{3\sqrt{n}}-\alpha_{n}-\P_{n,0}\Big(\max_{1\le k\le n+\sqrt{n}}S_{n,k}\ge \lambda_{n}\Big)\bigg)\le -\frac{1}{n}\log\frac{1}{4\sqrt{n}}. \label{eqn:bound_lg}
\end{align}
Then, together with~(\refeq{eqn:nae:r1}),~(\refeq{eqn:nae:type2}) and~\eqref{eqn:bound_lg}, we have that for sufficiently large $n$,
\begin{align}
R_{1}-\gamma &\le\frac{1}{n}\log\frac{1}{\beta_{n}}\\
&\le  \frac{\E_{n,0}[S_{n,n}]}{n}+\gamma+\frac{1}{n}\log (4\sqrt{n})\\
&\le \frac{\E_{n,0}[S_{n,n}]}{n}+2\gamma. \label{eqn:bd_R1}
\end{align}
Similarly, we have that for sufficiently large $n$,
\begin{align}
R_{0}-\gamma&\le \frac{1}{n}\log\frac{1}{\alpha_{n}}\le \frac{\E_{n,1}[-S_{n,n}]}{n}+2\gamma.  \label{eqn:bd_R0}
\end{align}
As $\frac{1}{n}(\E_{n,1}[-S_{n,n}],\E_{n,0}[S_{n,n}])$ is in the convex hull of $\mathcal{C}$, letting $\gamma\to 0^{+}$, we have from~\eqref{eqn:bd_R1} and~\eqref{eqn:bd_R0} that 
\begin{align}
(R_{0},R_{1})\in \overline{\mathrm{Conv}(\mathcal{C})}.
\end{align}
This completes the converse to Theorem~\ref{non-adaptive} under the expectation constraints.

Finally, we prove the converse to Theorem~\ref{non-adaptive} under the  probabilistic constraints. Let $(R_{0},R_{1})$ be an achievable error exponent pair. Suppose that $\big\{\big(\cX_n,\{\mu_{n,k},d_{n,k}\}_{k=1}^{\infty},T_{n}\big)\big\}_{n=1}^{\infty}$ is a sequence of SQHTs with non-adaptive strategies satisfying~(\refeq{eqn:R0}),~(\refeq{eqn:R1}), and~(\refeq{eqn:prob_con}).  
For any fixed  $\gamma$ such that $0< \gamma< \min\{R_{0},R_{1}, (1-\eps)/2\}$, we have from Definition~\ref{def:ach_ee} that
\begin{align}
\P_{n,0}(T_{n}>n)&\le \eps+\gamma,\\
R_{0}-\gamma&\le \frac{1}{n}\log\frac{1}{\alpha_{n}},\label{eqn:nap:r0}\\
R_{1}-\gamma&\le  \frac{1}{n}\log\frac{1}{\beta_{n}},\label{eqn:nap:r1}
\end{align}
 for sufficiently large $n$.  Using similar arguments as in the derivation of~(\refeq{type2p1}) with $\log\lambda_n:=\E_{n,0}[S_{n,n}]+n\gamma$, we have that
\begin{align}\label{eqn:nap:type2}
\log\frac{1}{\beta_n}
&\le \log\lambda_n-\log\bigg[1-\alpha_n-\P_{n,0}\bigg(\max_{1\le k\le n}S_{n,k}\ge \E_{n,0}[S_{n,n}]+n\gamma \bigg)-\eps-\gamma\bigg].
\end{align}
Then from~(\refeq{im1}), we have that
\begin{align}
\P_{n,0}\bigg(\max_{1\le k\le n}S_{n,k}\ge \E_{n,0}[S_{n,n}]+n\gamma \bigg)&\le  \frac{\E_{n,0}[(S_{n,n}-\E_{n,0}[S_{n,n}])^{2}]}{\gamma^{2}n^{2}}\\
&\le \frac{1}{\gamma^{2}n^{2}}\sum_{j=1}^{n}\E_{n,0}[(Z_{n,j}-\E_{n,0}[Z_{n,j}])^{2}]\\
&\le \frac{nC^{2}}{\gamma^{2}n^{2}},\label{eqn:nap:ptail}
\end{align}
where~(\refeq{eqn:nap:ptail}) follows from  Lemma~\ref{lemma:mre}, namely that $|Z_{n,j}|\le C$. Combining~(\refeq{eqn:nap:r0}), (\refeq{eqn:nap:type2}) and~(\refeq{eqn:nap:ptail}), we have that for sufficiently large $n$,
\begin{align}
R_{1}-\gamma\le \frac{1}{n}\log\frac{1}{\beta_n}
&\le \frac{1}{n} \E_{n,0}[S_{n,n}]+2\gamma.
\end{align}
Similarly we have 
\begin{align}
R_{0}-\gamma\le\frac{1}{n}\log\frac{1}{\alpha_n}
&\le \frac{ 1}{n}\E_{n,1}[-S_{n,n}]+2\gamma.
\end{align}
As $\frac{1}{n}(\E_{n,1}[-S_{n,n}],\E_{n,0}[S_{n,n}])$ is in the convex hull of $\mathcal{C}$, we have that
\begin{align}
\big(R_{0}-3\gamma,R_{1}-3\gamma\big)\in \overline{\mathrm{Conv}(\mathcal{C})}.
\end{align}
Letting $\gamma\to 0^{+}$, we have that 
\begin{align}
(R_{0},R_{1})\in \overline{\mathrm{Conv}(\mathcal{C})},
\end{align}
as desired. Thus we complete the proof of the converse part of Theorem~\ref{non-adaptive}.
\subsection{Proof of Theorem~\ref{thm:card}}\label{subsec:thm:card}

Before presenting the proof, we recapitulate the definition of an {\em extreme POVM}. 
\begin{definition}
Given a finite set $\mathcal{X}$, a POVM $m\in\M_{\mathcal{X}}$ is an extreme POVM if it cannot written as the convex combination of any two POVMs from $\M_{\mathcal{X}}$.
\end{definition}
The following theorem from~\cite[Theorem~2.21, pp.~23]{Holevo} characterizes the cardinality of non-zero elements of an extreme POVM. 
\begin{thm}\label{thm:holevo}
Let $\mathcal{X}$ be a finite set and let $m\in\M_{\mathcal{X}}$ be an extreme POVM. Then
\begin{align}
|\{x\in\mathcal{X}:m(x)\not=0\}|\le d^{2}.
\end{align}
\end{thm}
Now we proceed to prove Theorem~\ref{thm:card}. For any any two non-negative real numbers $t_{0}$ and $t_{1}$, let
\begin{align}
f(m;t_{0},t_{1}):=t_{0}D(P_{\rho_1,m}\|P_{\rho_0, m})+t_{1}D(P_{\rho_0,m}\|P_{\rho_1, m})
\end{align} 
and 
\begin{align}\label{eqn:maximum}
g(t_{0},t_{1}):=\sup_{\mathcal{X}}\max_{m\in\M_{\mathcal{X}}}f(m;t_{0},t_{1}),
\end{align}
where $\mathcal{X}$ runs over all finite sets and $m$ runs over all POVMs in $\mathcal{M}_{\mathcal{X}}$. It then follows from the supporting hyperplane theorem~\cite[Theorem 5, pp.~134]{Luenberger} that
\begin{align}
\overline{\mathrm{Conv}(\mathcal{C})}=\bigcap\limits_{(t_{0},t_{1}):\min\{t_{0},t_{1}\}>0}\left\{(R_{0},R_{1}):R_{0}\ge 0,R_{1}\ge 0,
t_{0}R_{0}+t_{1}R_{1}\le g(t_{0},t_{1})\right\}.
\end{align}
Therefore to prove Theorem~\ref{thm:card}, it suffices to show that
\begin{align}\label{eqn:rankone}
g(t_{0},t_{1})=\max_{m\in\M_{[d^{2}]}^{(1)}}f(m;t_{0},t_{1}).
\end{align}
For any finite set $\mathcal{X}$, let $\mathrm{Extr}(\M_{\mathcal{X}})$ be the set of extreme points of $\M_{\mathcal{X}}$. Since $f(m;t_{0},t_{1})$ is a convex function in $m$, then from~\cite[Corollary~32.3.1, pp.~344]{Rockafellar} it follows that
\begin{align}\label{eqn:extreme}
\max_{m\in\M_{\mathcal{X}}}f(m;t_{0},t_{1})=\max_{m\in\mathrm{Extr}(\M_{\mathcal{X}})}f(m;t_{0},t_{1}).
\end{align}
Let $m^{*}\in \mathrm{Extr}(\M_{\mathcal{X}})$ achieve the the maximum on the right-hand side of~(\refeq{eqn:extreme}). Then from Theorem~\ref{thm:holevo},  we obtain 
\begin{align}\label{eqn:cardub}
|\{x\in\mathcal{X}:m^{*}(x)\not=0\}|\le d^{2}.
\end{align}
Combining~(\refeq{eqn:extreme}) and~(\refeq{eqn:cardub}), we have that for any finite set $\mathcal{X}$,
\begin{align}\label{eqn:extremed2}
\max_{m\in\M_{\mathcal{X}}}f(m;t_{0},t_{1})\le \max_{m\in\mathrm{Extr}\big(\M_{[d^{2}]}\big)}f(m;t_{0},t_{1}),
\end{align}
which, together with~(\refeq{eqn:maximum}), further implies that
\begin{align}\label{eqn:extreme:d2}
g(t_{0},t_{1})=\sup_{\mathcal{X}}\max_{m\in\M_{\mathcal{X}}}f(m;t_{0},t_{1})\le \max_{m\in\mathrm{Extr}\big(\M_{[d^{2}]}\big)}f(m;t_{0},t_{1})\le g(t_{0},t_{1}).
\end{align}
Let $\bar{m}\in \mathrm{Extr}\big(\M_{[d^{2}]}\big)$ achieves the maximum on the right-hand side of~(\refeq{eqn:extremed2}). Then from~(\refeq{eqn:extreme:d2}) we have that
\begin{align}\label{eqn:maximum:d2}
g(t_{0},t_{1})=f(\bar{m};t_{0},t_{1})
\end{align}If $\bar{m}(x)$ is of rank one for all $x\in[d^{2}]$, the proof of~(\refeq{eqn:rankone}) is completed. Otherwise, for any $x\in[d^{2}]$, let $\sum_{k=1}^{r(x)}\bar{m}(x,k)$ be the spectral decomposition of $\bar{m}(x)$, where $\bar{m}(x,k)$ is a rank one matrix and $r(x)$ is the rank of $\bar{m}(x)$. Let  $\hat{m}:=\{\bar{m}(x,k):1\le k\le r(x), \ x\in[d^{2}]\}$. Then $\hat{m}$ is a POVM with outcomes taking values in $\{(x,k): 1\le k\le r(x), \ x\in[d^{2}]\}$.  It follows from Carath\'eodory's theorem~\cite[Theorem 17.1, pp.~155]{Rockafellar} that there exist $\hat{k}$ extreme POVMs $\{m_{j}\}_{j=1}^{\hat{k}}$ over $\{(x,k): 1\le k\le r(x), \ x\in[d^{2}]\}$ such that $\hat{k}\le d^{5}+1$ and
\begin{align}
\bar{m}(x,k)=\sum_{j=1}^{\hat{k}}\hat{t}_{j}m_{j}(x,k),
\end{align}
where $\hat{t}_{j}\in (0,1)$ and $\sum_{j=1}^{\hat{k}}\hat{t}_j=1$. Applying the data-processing inequality to the classical relative entropy, we have that
\begin{align}\label{eqn:chain}
f(\bar{m};t_{0},t_{1})\le f(\hat{m};t_{0},t_{1})\le \sum_{j=1}^{\hat{k}}\hat{t}_{j}f(m_{j};t_{0},t_{1}),
\end{align}
which, together with the fact that $f(m_{j};t_{0},t_{1})\le f(\bar{m};t_{0},t_{1})$ for any $1\le j\le \hat{k}$, implies that
\begin{align}
f(m_{j};t_{0},t_{1})= f(\bar{m};t_{0},t_{1})
\end{align}
for any $1\le j\le \hat{k}$. As $m_{j}$ is an extreme POVM, it follows from  Theorem~\ref{thm:holevo} that
\begin{align}
|\{(x,k): 1\le k\le r(x),m_{j}(x,k)\not=0,  x\in[d^{2}]\}|\le d^{2}.
\end{align}
As $0\le \hat{t}_{j}m_{j}(x,k)\le \bar{m}(x,k)$ and $\bar{m}(x,k)$ is a rank one matrix,  it follows that $m_{j}(x,k)$ is also a rank one matrix. Therefore, we have that
\begin{align}
f(\bar{m};t_{0},t_{1})=f(m_{j};t_{0},t_{1})\le \max_{m\in\M_{[d^{2}]}^{(1)}}f(m;t_{0},t_{1})\le f(\bar{m};t_{0},t_{1}),
\end{align}
which implies that~(\refeq{eqn:rankone}) holds. This completes the proof of Theorem~\ref{thm:card}.

\subsection*{Acknowledgements}
YL and VYFT are supported by a Singapore National Research Foundation Fellowship (R-263-000-D-02-281). MT is supported by NUS startup grants (R-263-000-E32-133 and R-263-000-E32-731) and by the National Research Foundation, Prime Minister's Office, Singapore and the Ministry of Education, Singapore under the Research Centres of Excellence programme.
\bibliographystyle{spmpsci}
\bibliography{reference,reference_MT}

\end{document}